\pdfoutput=1

\documentclass[prd,showpacs,letterpaper,10pt,aps,notitlepage,nofootinbib,groupedaddress]{revtex4-2}
\usepackage{changepage}
\usepackage[utf8]{inputenc}
\usepackage{amsfonts}
\usepackage{amsmath}
\usepackage{graphicx}
\usepackage{hyperref}
\usepackage{placeins}
\usepackage{mathrsfs}
\usepackage{bm}
\usepackage[utf8]{inputenc}
\usepackage{multirow}
\usepackage{slashed}
\usepackage[export]{adjustbox}
\hypersetup{colorlinks, linkcolor = [rgb]{0,0.0,0.75}, citecolor = [rgb]{0,0.0,0.75}, urlcolor = [rgb]{0,0.0,0.75}}
\usepackage{adjustbox}

\newcommand{\nb}{\phantom{0}}
\newcommand{\wm}{\phantom{-}}

\newcommand{\ev}[1]{\ensuremath{\left\langle #1 %
                     \right\rangle}} 

\begin{document}
\title{\texorpdfstring{$\bm{\Xi_c \to \Xi}$}{Xic to Xi} form factors from lattice QCD with domain-wall quarks: \texorpdfstring{\\}{} A new piece in the puzzle of \texorpdfstring{$\bm{\Xi_c^0}$}{Xic(0)} decay rates}

\author{Callum Farrell}
\author{Stefan Meinel}
\affiliation{Department of Physics, University of Arizona, Tucson, AZ 85721, USA}

\begin{abstract}
We present a lattice-QCD determination of the vector and axial-vector form factors that describe the charm-baryon semileptonic decays $\Xi_c\to \Xi \ell^+ \nu_\ell$. The calculation uses a domain-wall action for the up, down, and strange quarks, and an anisotropic clover action for the charm quark. We use four ensembles of gauge-field configurations generated by the RBC and UKQCD collaborations, with lattice spacings between 0.111 and 0.073 fm and pion masses ranging from 430 to 230 MeV. We present Standard-Model predictions for the decay rates and branching fractions of $\Xi_c^0\to \Xi_c^-\ell^+ \nu_\ell$ and $\Xi_c^+\to\Xi_c^0\ell^+ \nu_\ell$ for $\ell=e,\mu$. In particular, we obtain $\Gamma(\Xi_c^0 \to \Xi^- e^+ \nu_e)/|V_{cs}|^2 = 0.2515(73)\text{ ps}^{-1}$ and $\mathcal{B}(\Xi_c^0 \to \Xi^- e^+ \nu_e) = 3.58(12)\:\%$. These values are higher than those predicted by a previous lattice calculation and substantially higher than the experimentally measured values, but consistent with expectations from approximate $SU(3)$ flavor symmetry.
\end{abstract}

\maketitle

\FloatBarrier
\section{Introduction}
\FloatBarrier
\label{sec:intro}
Substantial experimental progress has been made in the study of charm-baryon semileptonic decays over the last decade. Following the first measurements of the absolute branching fractions of $\Lambda_c\to\Lambda e^+\nu_e$ and $\Lambda_c\to\Lambda\mu^+\nu_\mu$ by BESIII, published in 2015 and 2016 \cite{BESIII:2015ysy,BESIII:2016ffj}, further data taking also allowed recent analyses of the $q^2$ and angular distributions of these decays \cite{BESIII:2022ysa,BESIII:2023vfi} and a detailed comparison with a lattice-QCD calculation \cite{Meinel:2016dqj}. There have also been first measurements of $\Lambda_c$ semileptonic decay rates to other final states \cite{BESIII:2018mug,BESIII:2022qaf,BESIII:2024mgg}, with results consistent with lattice-QCD predictions \cite{Meinel:2017ggx,Meinel:2021mdj,Meinel:2021grq}.

In this work, we consider the $SU(3)$ partner process $\Xi_c \to \Xi \ell^+\nu_\ell$, which was first observed by ARGUS ($\Xi_c^0 \to \Xi^- \ell^+\nu_\ell$) \cite{ARGUS:1992jnv} and CLEO ($\Xi_c^+ \to \Xi^0 \ell^+\nu_\ell$) \cite{CLEO:1994aud}.
 In 2021, both Belle and ALICE reported new measurements of the relative branching ratio $\mathcal{B}(\Xi_c^0 \rightarrow \Xi^- e^+ \nu_{e})/\mathcal{B}(\Xi_c^0 \rightarrow \Xi^- \pi^+)$ \cite{Belle:2021crz, ALICE:2021bli} with the results
\begin{align}
    \frac{\mathcal{B}(\Xi_c^0 \rightarrow \Xi^- e^+ \nu_{e})}{\mathcal{B}(\Xi_c^0 \rightarrow \Xi^- \pi^+)} =
    \begin{cases}
        0.730 \pm 0.021 \pm 0.033, \text{   Belle \cite{Belle:2021crz}}, \\
        1.38 \pm 0.14 \pm 0.22, \text{   ALICE \cite{ALICE:2021bli}}. 
    \end{cases}
\end{align}
When combined with a 2018 Belle measurement of the normalization mode,
\begin{align}
    \mathcal{B}_{\text{Belle}}(\Xi_c^0 \rightarrow \Xi^- \pi^+) =  (1.80 \pm 0.50 \pm 0.14)\% \text{   \cite{Belle:2018kzz}},
\end{align}
these measurements yield
\begin{align}
    \mathcal{B}_{\text{Belle}}(\Xi_c^0 \rightarrow \Xi^- e^+ \nu_{e}) =& (1.31 \pm 0.04 \pm 0.07 \pm 0.38 ) \%, \label{eq:BXic0eBelle} \\
    \mathcal{B}_{\text{ALICE}}(\Xi_c^0 \rightarrow \Xi^- e^+ \nu_{e}) =& (2.48 \pm 0.25 \pm 0.40 \pm 0.72) \%, \label{eq:BXic0eALICE}
\end{align} 
where the third uncertainty is inherited from the normalization mode. Belle also analyzed the antimuon mode and found a branching fraction consistent with the positron mode \cite{Belle:2021crz}. The above measurements were used as inputs in a fit of several modes by the Particle Data Group (PDG), which yielded a value of
\begin{align}
    \mathcal{B}_{\text{PDG}}(\Xi_c^0 \rightarrow \Xi^- e^+ \nu_{e}) = (1.05 \pm 0.20) \% \text{ \cite{ParticleDataGroup:2024cfk}.} \label{eq:BXic0ePDG}
\end{align}
As has been pointed out in Refs.~\cite{Zhang:2021oja,He:2021qnc,Geng:2022yxb}, the experimental results for $\mathcal{B}(\Xi_c^0 \rightarrow \Xi^- e^+ \nu_{e})$ are considerably lower than expected based on the approximate $SU(3)$ flavor-symmetry relation to the decay $\Lambda_c \rightarrow \Lambda e^+ \nu_{e}$. At leading order, one would expect
\begin{equation}
\mathcal{B}_{SU(3)}(\Xi_c^0 \rightarrow \Xi^- e^+ \nu_{e})=\displaystyle\frac{3}{2}\frac{\tau_{\Xi_c^0}}{\tau_{\Lambda_c}}\mathcal{B}(\Lambda_c \rightarrow \Lambda e^+ \nu_{e})\approx 4.0\:\%, \label{eq:BXic0SU3LcL}
\end{equation}
where we used the latest BESIII measurement of $\mathcal{B}(\Lambda_c \rightarrow \Lambda e^+ \nu_{e})$ \cite{BESIII:2022ysa} and the latest lifetime values from PDG \cite{ParticleDataGroup:2024cfk} to numerically evaluate the right-hand side. Very recently, BESIII also reported the first observation of the decay $\Lambda_c \rightarrow n e^+ \nu_{e}$ \cite{BESIII:2024mgg}. Combined with the ratio of CKM matrix elements from a global fit \cite{UTfit:2022hsi}, this provides an alternative leading-order $SU(3)$-symmetry prediction of 
 \begin{equation}
 \mathcal{B}_{SU(3)}(\Xi_c^0 \rightarrow \Xi^- e^+ \nu_{e})=\displaystyle\frac{|V_{cs}|^2}{|V_{cd}|^2}\frac{\tau_{\Xi_c^0}}{\tau_{\Lambda_c}}\mathcal{B}(\Lambda_c \rightarrow n e^+ \nu_{e})\approx 5.0\:\%, \label{eq:BXic0SU3Lcn}
 \end{equation}
again considerably higher than the experimental measurements of $\mathcal{B}(\Xi_c^0 \rightarrow \Xi^- e^+ \nu_{e})$.

Ref.~\cite{Geng:2022yxb} suggested that the unexpectedly large $SU(3)$ breaking in $\Xi_c^0\to\Xi^-e^+\nu_e$ could be explained by a large $\Xi_c - \Xi'_c$ mixing angle, but lattice calculations \cite{Brown:2014ena,Liu:2023feb,Liu:2023pwr} found the mixing angle to be very small, in accordance with expectations based on heavy-quark symmetry and flavor symmetry.

Table~\ref{tab:branching_ratios} summarizes recent Standard-Model predictions of the $\Xi_c^0\to\Xi^-e^+\nu_e$ branching ratio. Note that also the predictions  based on potential models, QCD sum rules, or lattice QCD, which do not make any assumptions about $SU(3)$ flavor symmetry, are higher than the average of the experimental measurements. Moreover, the predictions published before 2020 use an outdated experimental value of the lifetime $\tau_{\Xi_c^0}=(112^{+13}_{-10}) \text{ fs}$ \cite{ParticleDataGroup:2018ovx}. The world average of this lifetime was substantially changed through new precise measurements by LHCb published in 2019 and 2021 \cite{LHCb:2019ldj,LHCb:2021vll}. The average of the two LHCb measurements is $\tau_{\Xi^0_c}= (152.0 \pm 2.0) \text{ fs}$ \cite{LHCb:2021vll}, while the latest world average reported by PDG is now $\tau_{\Xi^0_c}=(150.4 \pm 2.8) \text{ fs}$ \cite{ParticleDataGroup:2024cfk}. Updating the older theory predictions accordingly would further increase the predicted $\mathcal{B}(\Xi_c^0 \rightarrow \Xi^- e^+ \nu_{e})$. The tensions between experiment and theory invite further detailed study of the $\Xi_c \to \Xi  \ell^+ \nu_\ell$ transition, in particular using lattice QCD.

\begin{table}[]
    \centering
    \begin{tabular}{l c c c }
        \hline\hline \\[-3.2ex]
        & Method & $\mathcal{B}(\Xi_c^0 \rightarrow \Xi^- e^+ \nu_{e})$ [\%] \\
        \hline \\ [-3.2ex]
        Zhang \textit{et al.}, 2021   \cite{Zhang:2021oja} & Lattice QCD  & $2.38 \pm 0.30 \pm 0.33 $  \\ 
        Zhao \textit{et al.}, 2021 \cite{Zhao:2021sje} & QCD Sum Rules & $1.83 \pm 0.45$ \\
        He \textit{et al.}, 2021 \cite{He:2021qnc} & Flavor SU(3)  & $4.10 \pm 0.46$ \\
        Geng \textit{et al.}, 2021  \cite{Geng:2020gjh} & Light-Front Quark Model  & $3.49 \pm 0.95$  \\ 
        Faustov and Galkin, 2019 \cite{Faustov:2019ddj} & Relativistic Quark Model* & $2.38$  \\  
        Geng \textit{et al.}, 2019 \cite{Geng:2019bfz} & Flavor  SU(3)* & $3.0 \pm 0.3$  \\
        Zhao, 2018 \cite{Zhao:2018zcb} & Light-Front Quark Model*  & $1.35$  \\
        Geng \textit{et al.}, 2018 \cite{Geng:2018plk} & Flavor SU(3)*  &  $4.87 \pm 1.74$ \\
        Geng \textit{et al.}, 2017 \cite{Geng:2017mxn} & Flavor SU(3)* & $3.0 \pm 0.5$  \\
        Azizi \textit{et al.}, 2011 \cite{Azizi:2011mw} & Light-Cone QCD Sum Rules* & $7.26\pm2.54$  \\
        Liu and Huang, 2010 \cite{Liu:2010bh} & QCD Sum Rules* & $2.4$  \\
        \hline\hline
         &  \\
    \end{tabular}
    \caption{Recent theoretical predictions of $\mathcal{B}(\Xi_c^0 \rightarrow \Xi^- e^+ \nu_{e})$. The calculations denoted with a (*) used an outdated value of  $\tau_{\Xi^0_c}$, as explained in the main text.}
    \label{tab:branching_ratios}
\end{table}

The lattice calculation in Ref.~\cite{Zhang:2021oja} used two ensembles of lattice gauge configurations with pion masses of 290 and 300 MeV, and a clover action for all of the fermions. In the following, we present a new lattice-QCD calculation of the $\Xi_c\to\Xi$ form factors, using four ensembles of gauge configurations with domain-wall up, down, and strange quarks, and using a three-parameter heavy-quark action for the valence charm quark (preliminary results from this work were previously shown in Ref.~\cite{Farrell:2023vnm}). We compute three-point functions for up to 15 different source-sink separations to achieve good control over excited-state contamination. Our data cover a wide range of pion masses and lattice spacings, allowing us to perform a combined chiral and continuum extrapolation of the form factors. We obtain precise Standard-Model predictions for the differential and integrated $\Xi_c\to \Xi \ell^+ \nu_\ell$ decay rates.

\FloatBarrier
\section{Form-Factor Definitions}
\FloatBarrier

We use the same helicity-based form factor definitions introduced in Ref.~\cite{Feldmann:2011xf} and previously utilized in Refs.~\cite{Detmold:2015aaa, Detmold:2016pkz, Meinel:2016dqj, Meinel:2017ggx}, with the decomposition of the hadronic matrix elements given by

\begin{align}
 \nonumber \langle \Xi(p^\prime,s^\prime) | \overline{s} \,\gamma^\mu\, c | \Xi_c(p,s) \rangle = & \:
 \overline{u}_\Xi(p^\prime,s^\prime) \bigg[ f_0(q^2)\: (m_{\Xi_c}-m_\Xi)\frac{q^\mu}{q^2} \\
 & \phantom{\overline{u}_\Xi \bigg[}+ f_+(q^2) \frac{m_{\Xi_c}+m_\Xi}{s_+}\left( p^\mu + p^{\prime \mu} - (m_{\Xi_c}^2-m_\Xi^2)\frac{q^\mu}{q^2}  \right) \\
 \nonumber & \phantom{\overline{u}_\Xi \bigg[}+ f_\perp(q^2) \left(\gamma^\mu - \frac{2m_\Xi}{s_+} p^\mu - \frac{2 m_{\Xi_c}}{s_+} p^{\prime \mu} \right) \bigg] u_{\Xi_c}(p,s),
\end{align}

\begin{align}
 \nonumber \langle \Xi(p^\prime,s^\prime) | \overline{s} \,\gamma^\mu\gamma_5\, c | \Xi_c(p,s) \rangle =&
 -\overline{u}_\Xi(p^\prime,s^\prime) \:\gamma_5 \bigg[ g_0(q^2)\: (m_{\Xi_c}+m_\Xi)\frac{q^\mu}{q^2} \\
 & \phantom{\overline{u}_\Xi \bigg[}+ g_+(q^2)\frac{m_{\Xi_c}-m_\Xi}{s_-}\left( p^\mu + p^{\prime \mu} - (m_{\Xi_c}^2-m_\Xi^2)\frac{q^\mu}{q^2}  \right) \\
 \nonumber & \phantom{\overline{u}_\Xi \bigg[}+ g_\perp(q^2) \left(\gamma^\mu + \frac{2m_\Xi}{s_-} p^\mu - \frac{2 m_{\Xi_c}}{s_-} p^{\prime \mu} \right) \bigg]  u_{\Xi_c}(p,s).
\end{align}
Here, $q=p-p'$, $s_\pm =(m_{\Xi_c} \pm m_\Xi)^2 -q^2$ and $\sigma^{\mu\nu}=\frac{i}{2}(\gamma^\mu\gamma^\nu-\gamma^\nu\gamma^\mu)$. This helicity-based decomposition of the form factors obeys the endpoint relations
\begin{eqnarray}
 f_0(0) &=& f_+(0), \label{eq:endptconst1}  \\
 g_0(0) &=& g_+(0), \label{eq:endptconst2} \\
 g_\perp(q^2_{\rm max}) &=& g_+(q^2_{\rm max}), \label{eq:endptconst3} 
\end{eqnarray}
with $q^2_{\rm{max}}=(m_{\Xi_c}-m_{\Xi})^2$. 

\FloatBarrier
\section{Lattice Actions and Parameters}
\FloatBarrier

\begin{table}
 \begin{tabular}{lccccccccccccc}
\hline\hline \\ [-2.8ex]
Label & $N_s^3\times N_t \times N_5$ & $\beta$  & $a$ [fm] & $2\pi/L$ [GeV] &  $am_{u,d}$ &  $m_\pi$ [GeV] & $am_{s}^{(\mathrm{sea})}$ 
& $am_{s}^{(\mathrm{val})}$ & $\wm a m_Q^{(c)}$ & $\nu^{(c)}$ & $c_{E,B}^{(c)}$  & $N_{\rm ex}$ & $N_{\rm sl}$ \\
\hline
C01  & $24^3\times64\times16$ & $2.13$   & $0.1106(3)$ & $0.4673(13)$ & $0.01\nb$  & $0.4312(13)$ & $0.04$      & $0.0323$  & $\wm0.1541$ & $1.2004$ & $1.8407$  & 283 & 2264  \\
C005 & $24^3\times64\times16$ & $2.13$   & $0.1106(3)$ & $0.4673(13)$ & $0.005$    & $0.3400(11)$ & $0.04$      & $0.0323$  & $\wm0.1541$ & $1.2004$ & $1.8407$  & 311 & 2488  \\
F004 & $32^3\times64\times16$ & $2.25$   & $0.0828(3)$ & $0.4680(17)$ & $0.004$    & $0.3030(12)$ & $0.03$      & $0.0248$  & $-0.05167$   & $1.1021$ & $1.4483$  & 251 & 2008  \\
F1M  & $48^3\times96\times12$ & $2.31$   & $0.0728(3)$ & $0.3545(13)$ & $0.002144$ & $0.2320(10)$ & $0.02144$   & $0.02217$ & $-0.05874$  & $1.0941$ & $1.5345$ & 113 & 1808 \\
\hline\hline
\end{tabular}
\caption{Parameters of the four data sets used in this calculation. The generation of the ensembles and their lattice spacings are discussed in Refs.~\cite{RBC:2010qam,RBC:2014ntl,Boyle:2018knm}. For the charm quark action, the parameter values of the mass $a m_Q^{(c)}$, the anisotropy parameter $\nu^{(c)}$, and the chromoelectric/chromomagnetic clover coefficients  $c_E^{(c)}= c_B^{(c)}$ are discussed in Ref.~\cite{Meinel:2023wyg}. $N_{\rm ex}$ and $N_{\rm sl}$ are the number of exact and sloppy samples used in the all-mode-averaging procedure \cite{Blum:2012uh, Shintani:2014vja}. For all light-quark propagators, we use low-mode deflation. The sloppy samples are computed with reduced conjugate-gradient iteration counts.}
\label{tab:lattice_params}
\end{table}

We use four different ensembles of gauge-field configurations generated by the RBC and UKQCD collaborations with 2+1 flavors of domain-wall fermions and the Iwasaki gauge action  \cite{RBC:2010qam,RBC:2014ntl,Boyle:2018knm}. The main parameters of these ensembles and of the quark propagators we computed on them are listed in Table \ref{tab:lattice_params}. The ensembles we label as ``C01,'' ``C005,'' and ``F004'' \cite{RBC:2010qam,RBC:2014ntl} were generated with the Shamir domain-wall action, while ``F1M'' \cite{Boyle:2018knm} uses a M\"obius domain-wall action. For each ensemble, we implement the light and strange valence quarks with the same action as the sea quarks, with the valence light-quark mass set equal to the sea-quark mass. We use valence strange masses tuned to the physical point, which are slightly lower than the sea strange mass in the case of the C01, C005, and F004 ensembles, and slightly heavier in the case of the F1M ensemble.

For the charm quark, we use an anisotropic clover action of the same form as in Ref.~\cite{RBC:2012pds} (our notation for the bare parameters follows Ref.~\cite{Brown:2014ena}, while Ref.~\cite{RBC:2012pds} uses $m_0=m_Q$, $\zeta=\nu$, $c_P=c_E=c_B$). In contrast to Refs.~\cite{Detmold:2015aaa,Meinel:2016dqj,Meinel:2017ggx}, in which a two-parameter tuning based on the charmonium dispersion relation in combination with perturbation theory for $c_{E}^{(c)}$ and $c_{E}^{(c)}$ was used, here we employ the nonperturbative tuning of the three parameters $a m_Q^{(c)}$, $\nu^{(c)}$, and $c_{E}^{(c)}=c_{B}^{(c)}$ described in detail in Ref.~\cite{Meinel:2023wyg}. With this scheme, the tuned parameters replicate the experimental values the $D_s$ meson rest mass, kinetic mass, and hyperfine splitting \cite{Meinel:2023wyg}.

For the $c \to s$ currents, we use the mostly nonperturbative renormalization scheme of Refs.~\cite{Hashimoto:1999yp, El-Khadra:2001wco}. The forms of the currents are given explicitly in Eqs. (18)-(21) of  Ref.~\cite{Detmold:2015aaa}, but are generally structured as 
\begin{equation}
    J_\Gamma = \rho_\Gamma \sqrt{Z^{ll}_{V}Z^{cc}_{V}}\bigg[\bar{s}\Gamma c + \mathcal{O}(a) \text{ improvement terms} \bigg] \label{eq:current}.
\end{equation}
The majority of the renormalization is contained in the factors $Z^{ll}_{V}$ and $Z^{cc}_{V}$, the matching factors for the temporal components of the light-to-light and charm-to-charm vector currents. These are calculated nonperturbatively using charge conservation, with the specific values used in this work listed in Table \ref{tab:ZV}. The residual matching factors $\rho_\Gamma$ and the $\mathcal{O}(a)$-improvement terms used in this work are given in Table \ref{tab:Pmatchingfactors}. For the C005/C01 and F004 ensembles, they were computed by C.~Lehner at one loop in mean-field-improved lattice perturbation theory \cite{Lehner:2012bt}, for the slightly different bare charm masses as tuned in Ref.~\cite{Brown:2014ena}. For these two ensembles, the uncertainties given in Table \ref{tab:Pmatchingfactors} correspond to the change in the central value when changing the strong coupling from a mean-field lattice $\overline{\rm MS}$ coupling $\alpha_{s,{\rm lat}}^{\overline{\rm MS}}(a^{-1})$ to the continuum $\overline{\rm MS}$ coupling $\alpha_{s,{\rm ctm}}^{\overline{\rm MS}}(a^{-1})$ \cite{Flynn:2023nhi}. Due to the extreme smallness of the one-loop coefficients, the uncertainties of $\rho_{\rm \Gamma}$ may be underestimated, and we ultimately include an additional 1\% systematic uncertainty in the form factors to account for missing higher-order corrections in $\rho_{\rm \Gamma}$, as discussed in Sec.~\ref{sec:extrap}. The slight change in the bare charm masses from Ref.~\cite{Brown:2014ena} to Ref.~\cite{Meinel:2023wyg} is expected to have negligible effect, considering how little the coefficients change when going from the coarse to the fine lattice.

For the newer F1M ensemble, perturbative results were not available, and therefore we estimated the coefficients through an extrapolation of the values from the C005/C01 and F004 ensembles, done linearly in the lattice spacing. In this case, the uncertainties given in the table are the sum (in quadrature) of the F004 uncertainties and the shifts in the central values between F004 and F1M. Doing the extrapolation of the coefficients quadratically instead of linearly in the lattice spacing changes the values much less than the estimated uncertainties.

\begin{table}
\begin{center}
\small
\begin{tabular}{lllllllll}
\hline\hline \\[-2.5ex]
Ensemble          & \hspace{2ex} & $Z_V^{(ll)}$  & \hspace{2ex} & $Z_V^{(cc)}$     \\
\hline
  C005, C01     &&  $0.71273(26)$ \cite{RBC:2014ntl}        &&  $1.35761(16)$ \cite{Meinel:2023wyg}      \\[0.2ex]
  F004          &&  $0.7440(18)$  \cite{RBC:2014ntl}        &&  $1.160978(74)$ \cite{Meinel:2023wyg}      \\[0.2ex]
  F1M           &&  $0.7639(42)$  \cite{Marshall:2024pfg}   &&  $1.112316(61)$ \cite{Meinel:2023wyg}     \\[0.2ex]
\hline\hline
\end{tabular}
\caption{\label{tab:ZV} Nonperturbative results for the renormalization factors of the temporal components of the flavor-conserving vector currents.}
\end{center}
\end{table}

\begin{table}
\begin{center}
\small
\begin{tabular}{clllllll}
\hline\hline
Parameter         & \hspace{2ex}            & \hspace{1ex}  C005, C01     & \hspace{2ex} & \hspace{1ex} F004   & \hspace{2ex} &  \hspace{1ex} F1M   \\
\hline
 $\rho_{V^0}=\rho_{A^0}$     & & $\wm1.0027(11)$      & &  $\wm1.00195(59)$    & &  $\wm1.00152(73)$      \\[0.5ex] 
 $\rho_{V^j}=\rho_{A^j}$     & & $\wm0.9948(20)$      & &  $\wm 0.99675(99)$    & &  $\wm0.9978(15)$       \\[0.5ex] 
 $c_{V^0}^R=c_{A^0}^R$       & & $\wm0.0402(72)$       & &  $\wm 0.0353(53)$     & &  $\wm0.0326(60)$       \\[0.5ex] 
 $c_{V^0}^L=c_{A^0}^L$       & & $-0.0048(19)$          & &  $-0.00270(82)$        & &  $-0.0016(14)$         \\[0.5ex] 
 $c_{V^j}^R=c_{A^j}^R$       & & $\wm0.0346(50)$       & &  $\wm0.0283(32)$      & &  $\wm0.0249(47)$       \\[0.5ex] 
 $c_{V^j}^L=c_{A^j}^L$       & & $\wm0.00012(23)$      & &  $\wm0.00040(12)$     & &  $\wm0.00055(19)$      \\[0.5ex] 
 $d_{V^j}^R=-d_{A^j}^R$      & & $-0.0041(16)$         & &  $-0.0039(12)$        & &  $-0.0038(12)$         \\[0.5ex] 
 $d_{V^j}^L=-d_{A^j}^L$      & & $\wm0.00210(82)$          & &  $\wm0.00260(79)$      & &  $\wm0.00287(84)$      \\[0.5ex] 
\hline\hline
\end{tabular}\vspace{-2ex}
\end{center}
\caption{\label{tab:Pmatchingfactors} Perturbative renormalization and $\mathcal{O}(a)$-improvement coefficients for the $c\to s$ currents used in Eq.~(\ref{eq:current}). See the main text for details on their determination and uncertainties.} 
\end{table}

\FloatBarrier
\section{Correlation Functions and Ratios}
\FloatBarrier

\begin{table}
	\begin{tabular}{lccccccccc} \hline \hline 
		Ensemble   & \multicolumn{2}{c}{Up and down quarks} & \hspace{2ex} & \multicolumn{2}{c}{Strange quarks} & \hspace{2ex} & \multicolumn{2}{c}{Charm quarks} \\
               & $N_\textrm{Gauss}$ & $\sigma_\textrm{Gauss}/a$ & \hspace{2ex}  &  $N_\textrm{Gauss}$ & $\sigma_\textrm{Gauss}/a$ &  \hspace{2ex} & $N_\textrm{Gauss}$ & $\sigma_\textrm{Gauss}/a$  \\ \hline
    C005, C01  & $30$ & $4.350$ & \hspace{2ex}  & $30$ & $4.350$ & \hspace{2ex}  & $20$ & $3.0$ \\
    F004 & $60$ & $5.728$           & \hspace{2ex}  & $60$ & $5.728$ & \hspace{2ex}  & $35$ & $4.0$ \\ 
    F1M & $130$ & $8.9$           & \hspace{2ex}  & $70$ & $6.6$ & \hspace{2ex}  & $35$ & $4.5$ \\ \hline \hline
	\end{tabular}
	\caption{\label{tab:smearingparams}Parameters for the smearing of the quark fields in the baryon interpolating fields. The Gaussian smearing is defined as in Eq.~(8) of Ref.~\cite{Leskovec:2019ioa}. For the up, down, and strange quarks, we used APE-smeared gauge links \cite{Bonnet:2000dc} with $N_{\rm APE}=25$ and $\alpha_{\rm{APE}}=2.5$ in the Gaussian smearing kernel. For the charm quarks, we use Stout-smeared \cite{Morningstar:2003gk} gauge links with $N_{\rm Stout}=10$ and $\rho_{\rm Stout}=0.08$ in the Gaussian smearing.}
\end{table}

For the $\Xi_c$ and $\Xi$ baryons, we use the interpolating fields
\begin{align}
    \Xi_{c{\alpha}} = \epsilon_{abc} (C\gamma_5)_{\beta\gamma} u^a_\beta s^b_\gamma c^c_\alpha, \hspace{1em}
    \Xi_{\alpha} = \epsilon_{abc} (C\gamma_5)_{\beta\gamma} u^a_\beta s^b_\gamma s^c_\alpha, 
\end{align}
where the quark fields are smeared with the parameters given in Table \ref{tab:smearingparams}. We compute the forward and backward two-point correlation functions
\begin{equation}
\begin{aligned}
        C^{(2,\Xi,\text{fw})}_{\delta \alpha}(\textbf{p}',t) &= \sum_{\textbf{y}} e^{-i\textbf{p}' \cdot (\textbf{y}-\textbf{x})} \big<\Xi_{\delta}(x_0+t, \textbf{y}) \: \overline{\Xi}_\alpha(x_0,\textbf{x}) \big>, \\
        C^{(2,\Xi,\text{bw})}_{\delta \alpha}(\textbf{p}',t) &= \sum_{\textbf{y}} e^{-i\textbf{p}' \cdot (\textbf{x}-\textbf{y})} \big<\Xi_{\delta}(x_0, \textbf{x}) \: \overline{\Xi}_\alpha(x_0-t,\textbf{y}) \big>,\\
        C^{(2,\Xi_c,\text{fw})}_{\delta \alpha}(t) &= \sum_{\textbf{y}} \big<\Xi_{c\delta}(x_0+t, \textbf{y}) \: \overline{\Xi}_{c\alpha}(x_0,\textbf{x}) \big>, \\
        C^{(2,\Xi_c,\text{bw})}_{\delta \alpha}(t) &= \sum_{\textbf{y}} \big<\Xi_{c\delta}(x_0, \textbf{x}) \: \overline{\Xi}_{c\alpha}(x_0-t,\textbf{y}) \big>,
    \end{aligned}
\end{equation}
as well as the forward and backward three-point correlation functions
\begin{equation}
    \begin{aligned}
        C^{(3,\text{fw})}_{\delta \alpha}(\Gamma,\mathbf{p}',t,t') = \sum_{\mathbf{y},\mathbf{z}} e^{-i\mathbf{p}' \cdot (\mathbf{x}-\mathbf{y})} \Big<\Xi_{\delta}(x_0, \mathbf{x}) \:\: J_\Gamma(x_0 -t + t', \mathbf{y}) \:\:  \overline{\Xi}_{c\alpha}(x_0-t,\mathbf{z}) \Big>, \\
        C^{(3,\text{bw})}_{\delta \alpha}(\Gamma,\mathbf{p}',t,t-t') = \sum_{\mathbf{y},\mathbf{z}} e^{-i\mathbf{p}' \cdot (\mathbf{y}-\mathbf{x})} \Big<\Xi_{c\delta}(x_0 + t, \mathbf{z}) \:\: J^{\dagger}_\Gamma(x_0 + t', \mathbf{y}) \:\:  \overline{\Xi}_\alpha(x_0,\mathbf{x}) \Big>,
    \end{aligned}
\end{equation}
where $J_\Gamma$ is defined in Eq~(\ref{eq:current}). All correlation functions are computed using light and strange propagators with source position $(x_0,\mathbf{x})$, reusing propagators from Ref.~\cite{Meinel:2020owd} in the cases of C01, C005, F004 and from Ref.~\cite{Meinel:2023wyg} for F1M. In the three-point functions, the charm-quark propagators are sequential propagators.

\begin{figure}
    \vspace{-6ex}
    \centering
    \includegraphics[scale=.75]{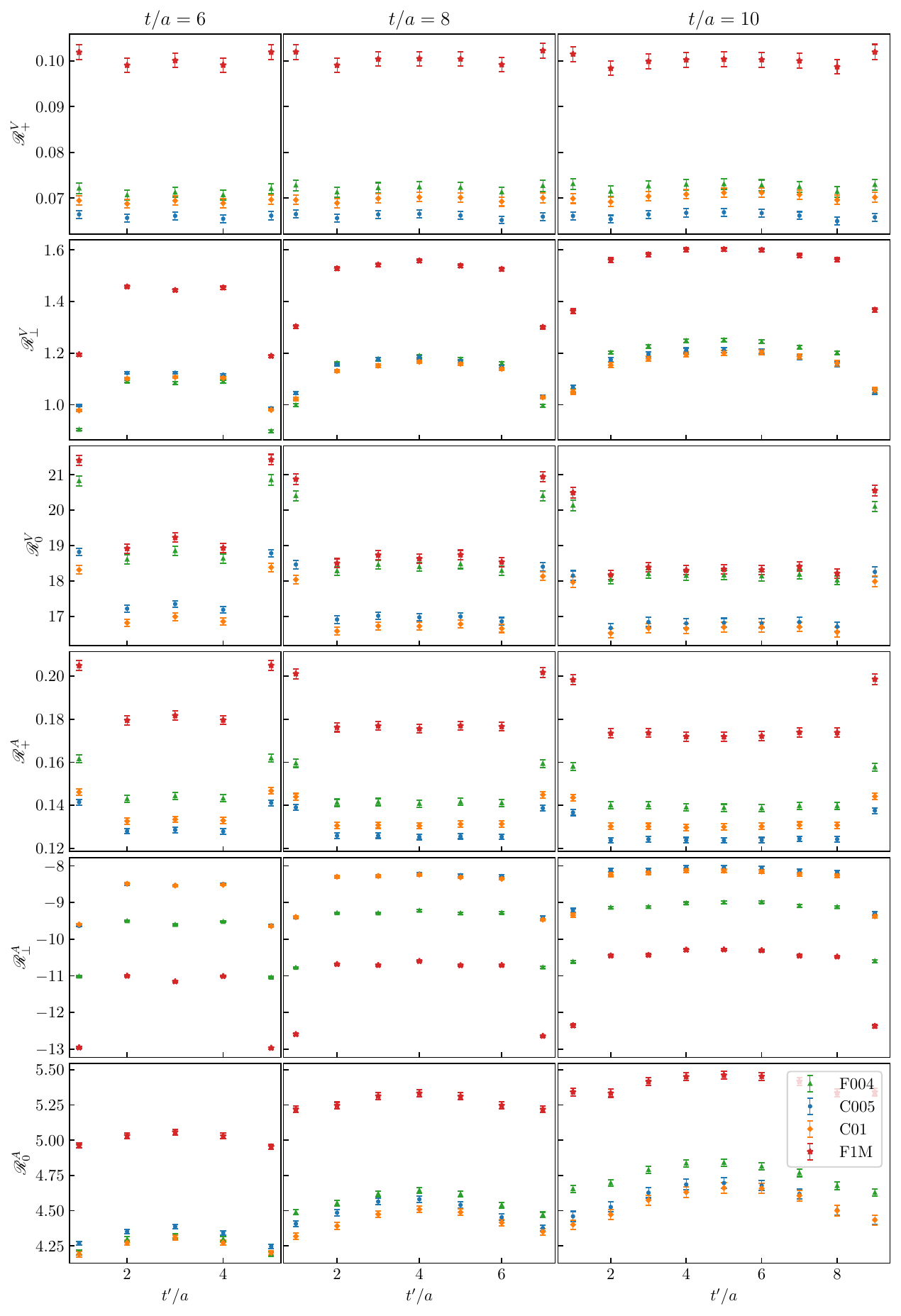}
    \caption{ The $t'$ dependence of both the vector-current and axial-vector-current ratios for three different values of the source-sink separation, $t$. The data from the C005, C01, and F004 ensembles are shown at $|\mathbf{p}^\prime|^2=1 \cdot (\tfrac{2\pi}{L})^2$, and the data from the F1M ensemble are shown at $|\mathbf{p}^\prime|^2=2\cdot (\tfrac{2\pi}{L})^2$. The plots are in units of $\rm{GeV}^{-2}$ for the dimensionful ratios $(\mathscr{R}^V_\perp, \mathscr{R}^V_0, \mathscr{R}^A_\perp, \mathscr{R}^A_0)$; the uncertainty from the lattice spacing is not shown. Note that the results from the different ensembles are not expected to be numerically close because of mass-dependent factors and because the $q^2$ values do not match exactly.}
    \label{fig:tpdep}
\end{figure}

As detailed in Ref.~\cite{Detmold:2015aaa}, the relevant helicity form factors can be obtained from ratios of these two-point and three-point correlation functions. These ratios remove the dependence on the overlap factors and the time-dependence of the ground-state. For example, the $f_+$ form factor can be isolated from the ratio
\begin{align}
    \mathscr{R}_{+}^V(\mathbf{p}^\prime,t,t^\prime) &=r_\mu[(1,\mathbf{0})] \: r_\nu[(1,\mathbf{0})] \frac{ \mathrm{Tr}\Big[   C^{(3,{\rm fw})}(\mathbf{p^\prime},\:\gamma^\mu, t, t^\prime) \:    C^{(3,{\rm bw})}(\mathbf{p^\prime},\:\gamma^\nu, t, t-t^\prime)  \Big] } 
    {\mathrm{Tr}\Big[C^{(2,\Xi,{\rm av})}(\mathbf{p^\prime}, t)\Big] \mathrm{Tr}\Big[C^{(2,\Xi_c,{\rm av})}(t)\Big] }, 
    \label{ratio1}
\end{align}
projected to longitudinal helicity with the virtual polarization vector $r_\mu[(1,\mathbf{0})]$, where
\begin{align}
    r[n]= n- \frac{(q \cdot n)}{q^2}q.
\end{align}
In the denominator of Eq.~(\ref{ratio1}), we use the averages of the forward and backward two-point functions. Sample plots of the numerical results for Eq. (17) and the other related ratios are shown in Fig.~\ref{fig:tpdep}. We then construct the quantity
\begin{align}
   \label{eq:Rfplus} R_{f_+}(|\mathbf{p}^\prime|, t)      =& \frac{2\, q^2 } {(E_\Xi-m_\Xi)(m_{\Xi_c}+m_\Xi)} \sqrt{\frac{ E_\Xi  }{ E_\Xi+m_\Xi } \mathscr{R}_{+}^V(|\mathbf{p}^\prime|, t, t/2)} \\
    =& \: f_+ + ({\rm excited\text{-}state\:\:contributions}), \nonumber
\end{align}
which approaches the value of the form factor at large Euclidean times when the excited-state contamination is suppressed. In Eq.~(\ref{eq:Rfplus}), the directions of $\mathbf{p}^\prime$ are averaged for a given value of $|\mathbf{p}^\prime|$, and we remove the dependence on the current insertion time $t'$ by setting $t'=t/2$ (or averaging over the two mid-points for odd $t/a$), minimizing excited state contamination for the given $t$. The values of the $\Xi_c$ and $\Xi$ baryon masses used above were determined by exponential fits of two-point correlation functions on each of the ensembles, with the resulting masses given in lattice units in Table~\ref{tab:hadronmasses}. 

The explicit form of the ratios ($R_{f_+}$, $R_{f_\perp}$, $R_{f_0}$, $R_{g_+}$, $R_{g_\perp}$, and $R_{g_0}$) computed to extract the different form factors are given in Eqs. (46)-(60) of Ref.~\cite{Detmold:2015aaa}. These ratios were computed at source-sink separations of $t/a=4,5,...,15$ on the C01 and C005 ensembles, $t/a=5,6,...,15$ on F004 and $t/a=6,7,...,20$ on F1M. We take our initial state $\Xi_c$ to have zero spatial momentum, and compute the ratios for values of $\Xi$ spatial momenta squared from $\{1,2,..,5\} \cdot (\tfrac{2\pi}{L})^2$ on C005, C01, F004 and $\{1,2,..,6,8\} \cdot (\tfrac{2\pi}{L})^2$ on F1M.

\begin{table}
\begin{tabular}{ccccccccccc}
\hline\hline
Ensemble  & \hspace{1ex} & $a m_{\Xi_c}$ & \hspace{1ex} & $am_{\Xi}$  & \hspace{1ex} & $am_{D}$ & \hspace{1ex} & $am_{D^*}$ & \hspace{1ex} & $am_{K}$     \\
\hline
C01     &&  $1.4328(19)$  &&  $0.8097(17)$  &&  $1.0712(30)$  && $1.1561(10)$ &&  $0.3247(11)$  \\
C005    &&  $1.4153(17)$  &&  $0.7836(16)$  &&  $1.0613(30)$  && $1.1464(11)$ &&  $0.3082(11)$  \\
F004    &&  $1.0551(16)$  &&  $0.5840(13)$  &&  $0.7922(29)$  && $0.85792(81)$ &&  $0.2249(10)$  \\ 
F1M     &&  $0.9212(14)$  &&  $0.5010(12)$  &&  $0.69105(47)$ && $0.74925(53)$ &&  $0.19100(21)$  \\ 
\hline\hline
\end{tabular}
\caption{\label{tab:hadronmasses}Hadron masses in lattice units. The $D^{(*)}$ and $K$ masses are used in the $z$-expansion parametrization of the form factors.}
\end{table}

\FloatBarrier
\section{Extrapolations to Infinite Source-Sink Separation}
\label{sec:AIC}
\FloatBarrier

Isolating the ground-state contribution in Eq.~(\ref{eq:Rfplus}) requires that the ratios be extrapolated to infinite source-sink separation. As in our previous calculations (Refs.~\cite{Detmold:2015aaa, Detmold:2016pkz, Meinel:2016dqj, Meinel:2017ggx}), we achieve this using the fit functions
\begin{align}
    R_{f,i,n}(t) = f_{i,n} + A_{f,i,n} \: e^{-\delta E_{f,i,n}\:t}, \hspace{2ex} \delta E_{f,i,n}=\delta E_{\rm min} + e^{\,l_{f,i,n}}\:\:{\rm GeV}
    \label{eq:ratiofitfunc}
\end{align}
with fit parameters $f_{i,n}$, $A_{f,i,n}$, and $l_{f,i,n}$. Here the index $f$ labels the different form factors, the index $n$ labels the different momenta, and the index $i$ labels the different ensembles. The $f_{i,n}$ term is the value of the form factor extrapolated to infinite source-sink separation, while the leading excited-state contamination is parameterized by the exponential term. Here we set the minimum energy gap $\delta E_{\rm min}$  to 100 MeV, while the energy gap parameters $l_{f,i,n}$  are constrained by priors (analogously to Eq.~(70) of Ref.~\cite{Detmold:2015aaa}) so as not to vary drastically across the different ensembles. Note that, because we set $t'=t/2$ in the three-point functions, $\delta E_{f,i,n}$ corresponds to only half the energy gap for the lowest excited-to-ground-state contribution in the three-point function.

As in the previous work \cite{Detmold:2015aaa, Detmold:2016pkz, Meinel:2016dqj, Meinel:2017ggx}, for a given momentum $n$, all of the vector ($f=f_+,f_\perp, f_0$) or all of the axial-vector ($f=g_+,g_\perp, g_0$) form factors are fit simultaneously. Additionally, both of these fits include data in an alternative basis for the form factors, the ``Weinberg basis,'' defined in Eqs.~(6,7) of Ref.~\cite{Detmold:2015aaa}. Linear combinations of the form factors in the Weinberg basis are related to the helicity form factors by Eqs.~(8-13) of Ref.~\cite{Detmold:2015aaa}. Performing a simultaneous fit of the $ R_{f,i,n}(t)$ including both form factor bases and enforcing that the extracted form factors $f_{i,n}$ obey Eqs.~(8)-(13) of Ref.~\cite{Detmold:2015aaa} significantly constrains and stabilizes the fits. These constraints are imposed as priors in the $\chi^2$ function, as in Eq.~(70) of Ref.~\cite{Detmold:2015aaa}. Furthermore, the data from the ensembles $i=\rm C005, C01, F004$ share common momenta, and are fit simultaneously, linked by priors on the differences of the energy gap parameters $l_{f,i,n}$. The data from the $\rm F1M$ ensemble are fit separately.

In previous work, systematic uncertainties associated with the choices of fit ranges for the $ R_{f,i,n}(t)$ were estimated from the shifts in $f_{i,n}$ when increasing $t_{\rm min}$ by one unit. Here, we improve on this strategy by adopting a Bayesian model-averaging approach proposed in Ref.~\cite{Jay:2020jkz} and further developed in Ref.~\cite{ Neil:2023pgt}. Specifically, we use the ``perfect model'' Akaike information criterion (AIC)  Refs.~\cite{Jay:2020jkz, Neil:2023pgt}. This criterion is used to average over fits with different ranges of $t$.

\begin{figure}
    \centering
    \includegraphics[width=\linewidth]{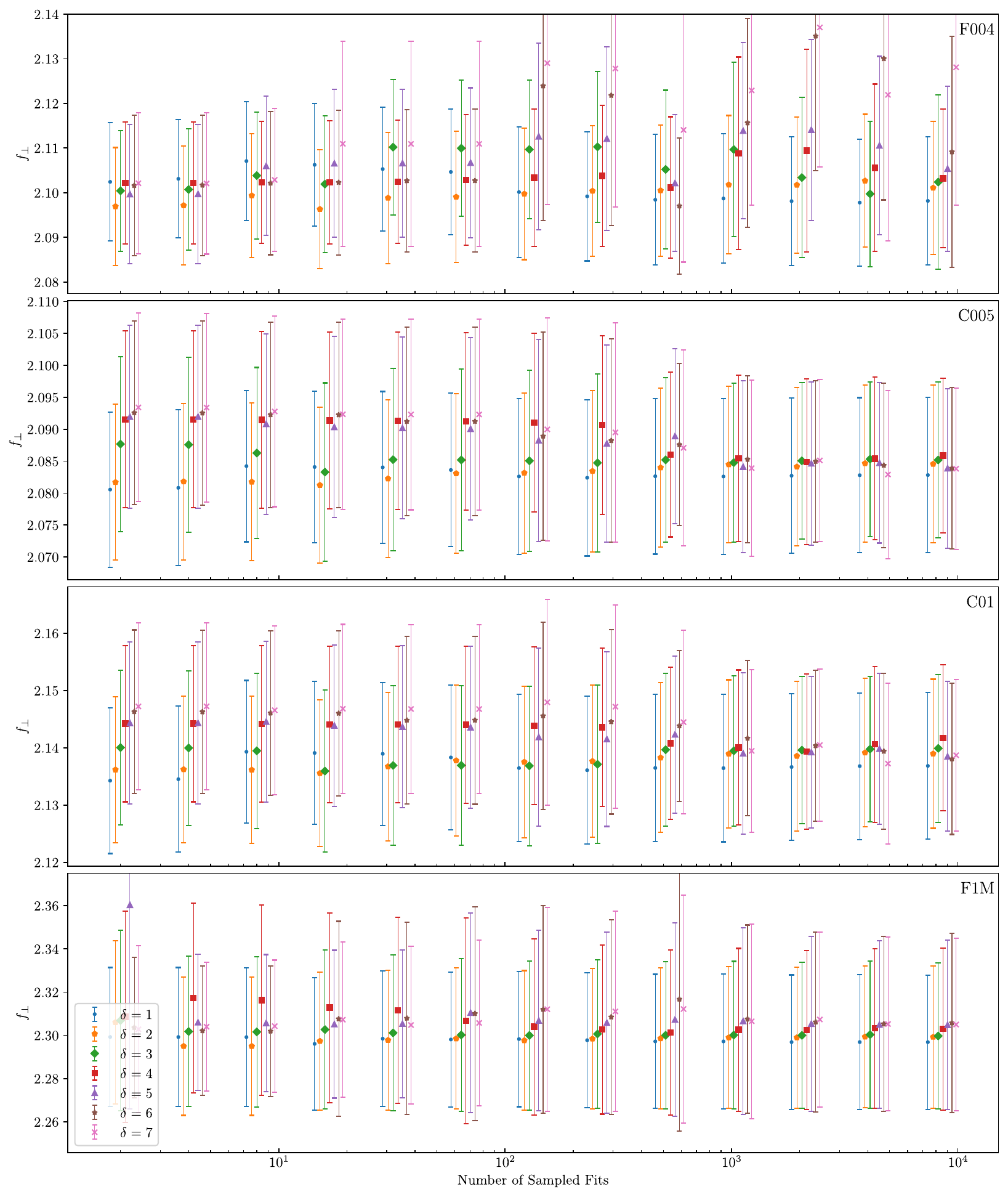}
    \caption{Evolution of the AIC average and uncertainty of the form factor $f_\perp$ at $|\mathbf{p}^\prime|^2=1\cdot (\tfrac{2\pi}{L})^2$ as a function of the number of sample fits, for widths of the $t_{\rm min}$ random distributions equal to $\delta=1,...,7$ (in lattice units). Values of $\delta > 4$ required larger numbers of sample fits to comprehensively explore the model space, but trended toward the same central value. The larger model space gained by increasing $\delta$ adds many models, but they have comparatively small model weights as the number of cut data points is increased without corresponding improvement in the $\chi^2$. Therefore, we use $\delta=4$ and 10,000 sample fits to obtain the final estimates.
    }
    \label{fig:delta_dep}
\end{figure}

As explained in Ref.~\cite{Jay:2020jkz}, the AIC average for a fit parameter (or a function of the fit parameters) is a weighted average 
\begin{equation}
    \ev{a_0} = \sum_M \ev{a_0}_M \mathrm{pr}(M|D) \label{eq:model_avg_mean},
\end{equation}
with the model weights
\begin{equation}
   \mathrm{pr}(M|D) = C \exp \left[ -\frac{1}{2} \left( \mathrm{AIC}^{\mathrm{perf}} \right) \right].
\end{equation}
Here,
\begin{equation}
   \mathrm{AIC}^{\mathrm{perf}} =  \chi_{M}^2 + 2k -2d_M, 
\end{equation}
where $\chi_{M}^2$ is the chi squared for a given model (fit) $M$, $k$ is the number of model parameters, and $d_M$ is the number of data points included in the specific fit $M$. The normalization factor $C$ ensures that $\Sigma_M\mathrm{pr}(M|D)=1$. The variance of the AIC average is computed as
\begin{equation}
\sigma^2_{a_0} = \sum_M \sigma^2_{a_0,M} \mathrm{pr}(M|D) + \sum_M \ev{a_0}_M^2 \mathrm{pr}(M|D) - \left( \sum_M \ev{a_0}_M \mathrm{pr}(M|D)   \right)^2. \label{eq:mw_var}
\end{equation}
Critically, the last two terms in this variance account for the systematic uncertainty associated with the choice of $t_{\rm min}$, eliminating the need for the previously utilized procedure. 

\begin{figure}
    \centering
    \includegraphics[width=\linewidth]{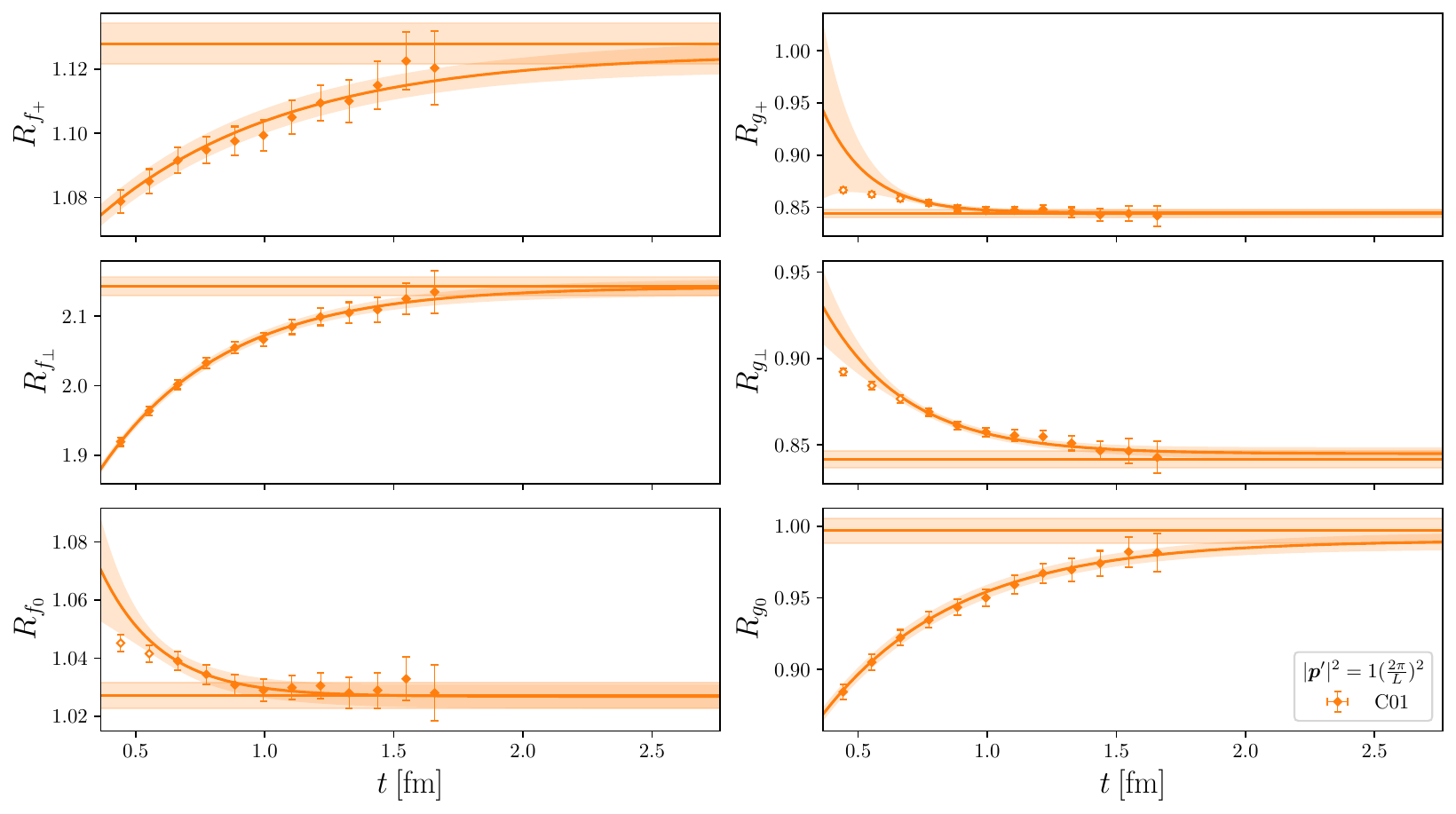}
    \caption{Plots showing the AIC analysis of the quantities $R_f(|\mathbf{p}^\prime|,t)$ for the $\rm{C01}$ ensemble at $|\mathbf{p}^\prime|^2=1(2\pi/L)^2$. The curves going through the data points belong to the sample fit with the highest model weight, with the bands showing only the statistical uncertainty, whereas the horizontal bands depict the AIC average values of the extracted ground-state form factors and their total uncertainty. Data points plotted with open symbols were omitted in the highest-model-weight fit.}   \label{fig:c01_ratios}
\end{figure}

The fits using Eq.~(\ref{eq:ratiofitfunc}) are performed simultaneously over all the vector or axial-vector form factors, and including both form factor bases. Therefore, for a single $R_{f,i,n}$, there are six individual form factors that must be fit simultaneously, and six corresponding individual values of $t_{\rm min}$ that must be varied.  For the three-ensemble fit, there are then 18 different values of $t_{\rm min}$ that must be varied over a wide enough range, with each combination of these $t_{\rm min}$'s contributing a unique model $M$ to the AIC average. To avoid performing a computationally infeasible number of fits, we randomly sample the high-dimensional space of $t_{\rm min}$'s. For every fit, we draw each $t_{\rm min}$ from a uniform random distribution ranging from the smallest value for which we have data to a value that is larger by some chosen $\delta$. The parameter $\delta$ therefore corresponds to the width of the random distribution of $t_{\rm min}$.

By performing a large enough number of these fits with randomly sampled $t_{\rm min}$, we can explore the model space densely enough to estimate the AIC average. To determine when a sufficiently thorough exploration of the model space is achieved, we investigated the behavior of the AIC average as a function of the value of $\delta$ and the number of randomly sampled fits, as shown for one of the form factors at one momentum in Fig.~\ref{fig:delta_dep}. 

As explained in the caption of Fig.~\ref{fig:delta_dep}, we set $\delta=4$ and performed 10,000 fits with randomly sampled values of $t_{\rm min}$ to obtain the final estimates. In all cases, we found that the AIC average values were consistent across choices of $\delta \geq 2$ and for a sufficiently large ($\geq$ 7,000) numbers of random fits. Furthermore, the results of the AIC averaging procedure were consistent with the fit results obtained using the procedure previously utilized in Refs.~\cite{Detmold:2015aaa, Detmold:2016pkz, Meinel:2016dqj, Meinel:2017ggx}, and had more stable estimates of the systematic uncertainty resulting from the choice of $t_{\rm min}$. However, for larger $q^2$ values, the terms quantifying this systematic uncertainty in Eq.~(\ref{eq:mw_var}) were drastically smaller than for smaller $q^2$ values. To be conservative, we enlarged the estimates of systematic uncertainties for these points through the prescription discussed in Section~\ref{sec:extrap}. 

Examples of the results of this fitting prescription are shown in Fig.~\ref{fig:c01_ratios}, in which the highest-weight fit to the ratio data from the $\rm{C01}$ ensemble at $|\mathbf{p}'|^2 = 1\cdot (\tfrac{2\pi}{L})^2$ is shown along with the AIC average value and uncertainty (the analogous plots for the other ensembles are given in Fig.~\ref{fig:other_ensemble_ratios}, in the appendix).

\FloatBarrier
\section{Chiral-continuum extrapolation}
\label{sec:extrap}
\FloatBarrier
We obtain parameterizations of these form factors in the physical limit ($a=0$, $m_{\pi} =m_{\pi, \rm{ phys}}$) by performing a BCL $z$-expansion \cite{Bourrely:2008za} for each of the form factors, with additional terms added to incorporate both the dependence on the pion mass and on the lattice spacing, as in Refs.~\cite{Detmold:2015aaa, Detmold:2016pkz, Meinel:2016dqj, Meinel:2017ggx}. After factoring out the expected $D_s$ meson poles, each form factor $f$ is expressed through an expansion in the variable
\begin{equation}
z_f(q^2) = \frac{\sqrt{t^f_+-q^2}-\sqrt{t^f_+-t_0}}{\sqrt{t^f_+-q^2}+\sqrt{t^f_+-t_0}}.
\end{equation}
This transformation maps the complex $q^2$ plane onto the unit disk, with the parameter
\begin{equation}
t_0 = q^2_{\rm max} = (m_{\Xi_c} - m_{\Xi})^2
\end{equation}
defining which particular value of $q^2$ is mapped to the center of the unit disk. The parameters $t_{+}^f$ are set to the positions at which the two-particle branch cuts along the real $q^2$ axis begin. Since our calculation is performed in the limit of exact isospin symmetry, we use
\begin{align}
\nonumber t^{f_+,f_\perp,f_0}_+ &= (m_D + m_K)^2, \\
t^{g_+,g_\perp,g_0}_+ &= (m_{D^*} + m_K)^2. \label{eq:tplus}
\end{align}
We fit the lattice data with the functions
\begin{eqnarray}
\nonumber f(q^2) &=& \frac{1}{1-(a^2 q^2)/(a m^f_{\rm pole})^2}
\bigg[ a_0^f\bigg(1+c_0^f \frac{m_\pi^2-m_{\pi,{\rm phys}}^2}{\Lambda_\chi^2}+\widetilde{c}_0^f \frac{m_\pi^3-m_{\pi,{\rm phys}}^3}{\Lambda_\chi^3}\bigg) \\
&& \hspace{30ex} +\: a_1^f\bigg(1+c_1^f\frac{m_\pi^2-m_{\pi,{\rm phys}}^2}{\Lambda_\chi^2}\bigg)\:z_f(q^2)  + a_2^f\:z_f^2(q^2)  + a_3^f\:z_f^3(q^2) \bigg] \\
\nonumber && \times\: \bigg[1  + b^f\, a^2|\mathbf{p^\prime}|^2 + d^f\, a^2\Lambda_{\rm had}^2
                     + \widetilde{b}^f\, a^4 |\mathbf{p^\prime}|^4
                     + \widehat{d}^f\, a^3 \Lambda_{\rm had}^3
                     + \widetilde{d}^f a^4 \Lambda_{\rm had}^4
                     + j^f   a^4 |\mathbf{p^\prime}|^2\Lambda_{\rm had}^2 \bigg], \hspace{5ex}  \label{eq:FFfit}
\end{eqnarray}
which in the physical limit simplify to the form
\begin{equation}
 f(q^2) = \frac{1}{1-q^2/(m_{\rm pole}^f)^2} \big[ a_0^f + a_1^f\:z_f(q^2) + a_2^f\:z_f^2(q^2)  + a_3^f\:z_f^3(q^2) \big].\label{eq:zphys}
\end{equation}
The physical values of the pole masses and their quantum numbers are listed in Table~\ref{tab:polemasses}. We use the lattice values of $am_{\Xi_c}$, $a m_{\Xi}$, $a m_D^{(*)}$, and $a m_K$ computed on each ensemble to evaluate $a^2 q^2$ and $z$ when fitting Eq.~(\ref{eq:FFfit}). We eliminate the parameters $a_2^{f_0,g_0}$ with the helicity endpoint constraints in Eqs.~(\ref{eq:endptconst1})-(\ref{eq:endptconst2}), and incorporate Eq.~(\ref{eq:endptconst3}) by using the single parameter $a^{g_+,g_\perp}_0$ for both the $g_+$ and $g_\perp$ form factors. The scale factors $\Lambda_{\rm had} = 300$ MeV and $\Lambda_\chi = 4\pi f_\pi$ (with $f_\pi =132$ MeV) allow all parameters to remain dimensionless.

Unlike in previous work \cite{Detmold:2015aaa, Detmold:2016pkz, Meinel:2016dqj, Meinel:2017ggx}, here we do not perform separate ``nominal'' and ``higher-order'' fits, and instead directly use a ``higher-order'' fit to obtain the physical-limit form factors. This has the advantage that the resulting covariance matrix of the fit parameters directly gives the total (statistical plus systematic) uncertainties.

The parameters $a_0^f$, $a_1^f$, $a_2^f$, $c_0^f$, $b^f$, and $d^f$ were left unconstrained, while the higher-order parameters $a_3^f$, $\widetilde{c}_0^f$, $c_1^f$, $\widetilde{b}^f$, $\widehat{d}^f$, $\widetilde{d}^f$, and $j^f$, which are not needed to describe the data, were constrained with Gaussian priors with central values and widths as in Eqs.~(40)-(48) of Ref.~\cite{Detmold:2016pkz}, and $a_3^f=0 \pm 30$. By including these higher-order terms with priors limiting them to be not unnaturally large, systematic uncertainties from such higher-order effects are incorporated in the final form-factor results.

Besides the effects of the higher-order pion-mass and lattice-spacing terms, our fit also incorporates the following sources of systematic uncertainty:
\begin{enumerate}
    \item When renormalizing the vector and axial-vector currents, the bootstrap samples for the ratios (\textit{e.g.} Eq~(\ref{eq:Rfplus})) were generated with the residual matching factors and the $\mathcal{O}(a)$-improvement coefficients drawn from Gaussian random distributions, with central values and widths given in Table~\ref{tab:Pmatchingfactors}. Furthermore, in the data covariance matrix used for the fit, we included an additional $1\%$ renormalization uncertainty for both the vector and axial-vector form factors to more conservatively account for missing higher-order corrections to the residual matching factors $\rho_{\Gamma}$; this uncertainty is taken to be 100\% correlated between either the vector or axial-vector form factors.
    \item In Refs.~\cite{Detmold:2015aaa,Meinel:2017ggx}, the finite-volume errors in the $\Lambda_b \to N$ and $\Lambda_c \to N$ form factors were estimated to be 3\% for the smallest value of $m_\pi L$ used there. It is reasonable to expect similar behavior for the $\Xi_c\to\Xi$ form factors. However, the smallest value of $m_\pi L$ is larger here, and we therefore rescale the above estimate with the exponential of the ratio of the smallest $m_\pi L$, leading to an estimate of 1\%. The isospin breaking effects are again estimated to be $\mathcal{O}((m_d-m_u)/\Lambda_{\rm QCD}) \approx 0.5\%$, and $\mathcal{O(\alpha_{\rm e.m.)}} \approx 0.7\%$. The finite-volume and isospin-breaking uncertainties were added to the data covariance matrix used in the fit, assuming $100\%$ correlation between either the vector or axial-vector form factors.
    \item The uncertainties in the lattice spacings and pion masses were included by promoting these values to fit parameters with Gaussian priors determined by the central values and uncertainties listed in Table~\ref{tab:lattice_params}.
    \item In the AIC averages, the terms in Eq~(\ref{eq:mw_var}) describing the systematic uncertainty from the $t_{\rm min}$ variation were sometimes orders of magnitude smaller at high values of $q^2$ compared to the systematic uncertainties at low $q^2$. To be more conservative, for each form factor and ensemble, we computed the average of these systematic uncertainties across all $q^2$ values, and then replaced any estimates that fell below the average by the average.
\end{enumerate}

\begin{table}
\begin{tabular}{ccccc}
\hline\hline
 $f$     & \hspace{1ex} & $J^P$ & \hspace{1ex}  & $m_{\rm pole}^f$ [GeV]  \\
\hline
$f_+$, $f_\perp$,                                          && $1^-$   && $2.112$  \\
$f_0$                                                      && $0^+$   && $2.318$  \\
$g_+$, $g_\perp$,                                          && $1^+$   && $2.460$  \\
$g_0$                                                      && $0^-$   && $1.968$  \\
\hline\hline
\end{tabular}
\caption{\label{tab:polemasses}$D_s$ meson poles in each of the different form factors.}
\end{table}

\begin{table}
\begin{tabular}{cc}
\hline \hline
 Parameter & Value  \\
\hline
$a_0^{f_{+}}$  & $ \wm 0.957 \pm 0.020$ \\
$a_1^{f_{+}}$  & $-4.483 \pm 0.453$ \\
$a_2^{f_{+}}$  & $ \wm 16.22 \pm 5.170$ \\
$a_3^{f_{+}}$  & $-20.33 \pm 28.79$ \\
$a_0^{f_{\perp}}$  & $ \wm 1.757 \pm 0.042$ \\
$a_1^{f_{\perp}}$  & $-6.555 \pm 0.614$ \\
$a_2^{f_{\perp}}$  & $ \wm 12.56 \pm 6.072$ \\
$a_3^{f_{\perp}}$  & $-6.574 \pm 29.63$ \\
$a_0^{f_{0}}$  & $ \wm 0.903 \pm 0.020$ \\
$a_1^{f_{0}}$  & $-3.473 \pm 0.428$ \\
$a_3^{f_{0}}$  & $-4.710 \pm 28.72$ \\
$a_0^{g_{+},g_{\perp}}$  & $ \wm 0.764 \pm 0.014$ \\
$a_1^{g_{+}}$  & $-3.455 \pm 0.271$ \\
$a_2^{g_{+}}$  & $ \wm 12.73 \pm 4.217$ \\
$a_3^{g_{+}}$  & $ \wm 7.531 \pm 28.40$ \\
$a_1^{g_{\perp}}$  & $-3.295 \pm 0.336$ \\
$a_2^{g_{\perp}}$  & $ \wm 11.46 \pm 4.896$ \\
$a_3^{g_{\perp}}$  & $ \wm 13.89 \pm 29.26$ \\
$a_0^{g_{0}}$  & $ \wm 0.836 \pm 0.017$ \\
$a_1^{g_{0}}$  & $-5.143 \pm 0.380$ \\
$a_3^{g_{0}}$  & $ \wm 0.090 \pm 28.91$ \\
\hline\hline
\end{tabular}
\caption{\label{tab:zfit_params} Results for the $z$-expansion parameters needed to describe the form factors in the physical limit as shown in Eq.~(\ref{eq:zphys}).
Machine-readable files with the parameter values and their covariance matrix are provided in the supplemental material \cite{Supplemental}. The correlation matrix is also given in in Table~\ref{tab:corr} in the Appendix.}
\end{table}

\begin{figure}
    \centering
    \hspace{1cm}
    \includegraphics[width=0.8\linewidth]{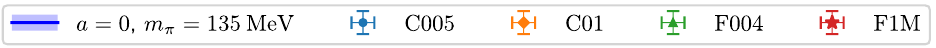}
    \includegraphics[width=\linewidth]{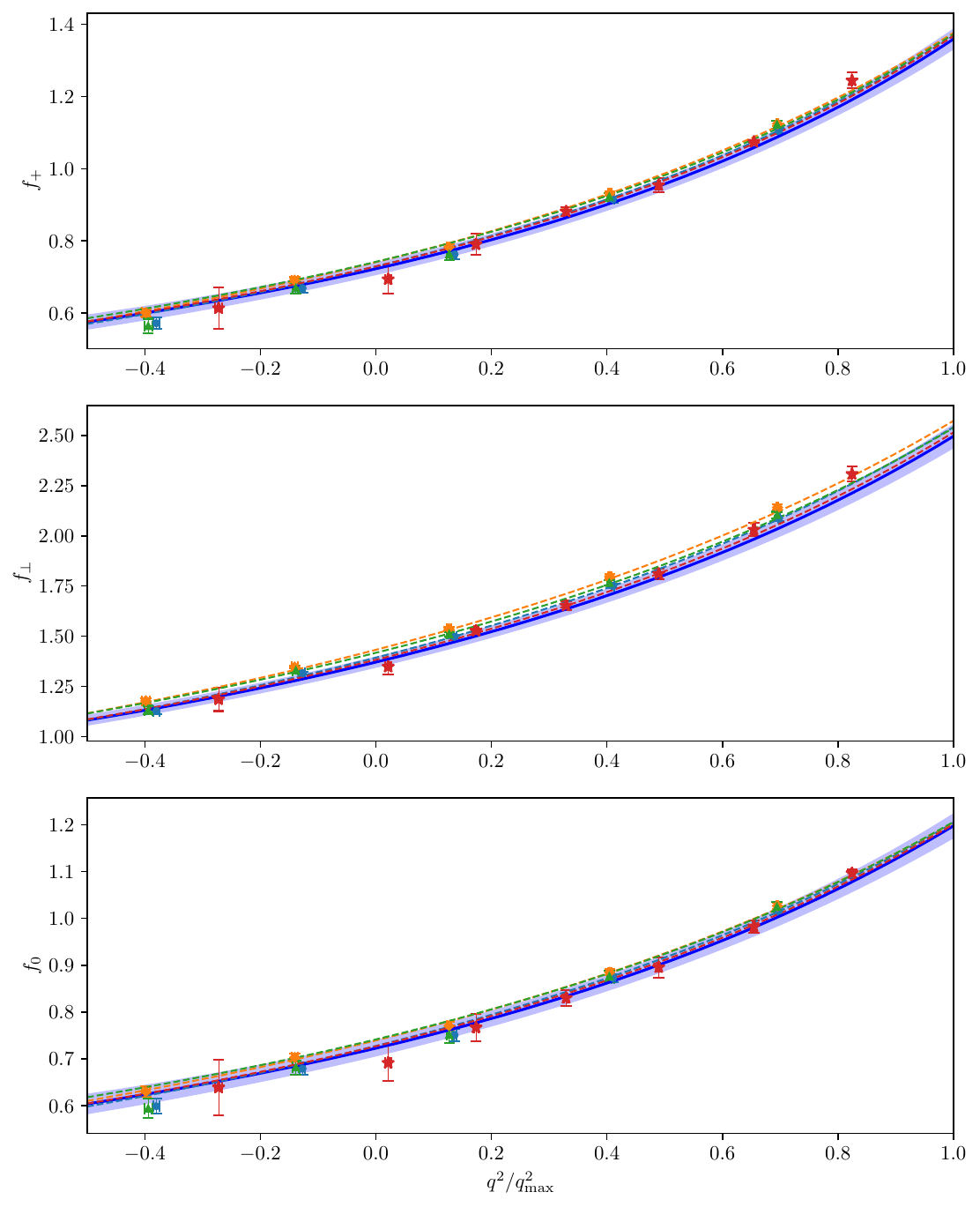}
    \caption{Chiral and continuum extrapolations of the $\Xi_c \to \Xi$ vector form factors. The solid blue lines show the form factor curves in the physical limit, while the dashed lines show the modified $z$-expansion fits evaluated with the individual lattice spacings and pion masses for each ensemble. The bands include the combined statistical and systematic uncertainties.}
    \label{fig:vector_extrap}
\end{figure}

\begin{figure}
    \centering
    \hspace{1cm}
    \includegraphics[width=0.8\linewidth]{ff_legend.pdf}
    \includegraphics[width=\linewidth]{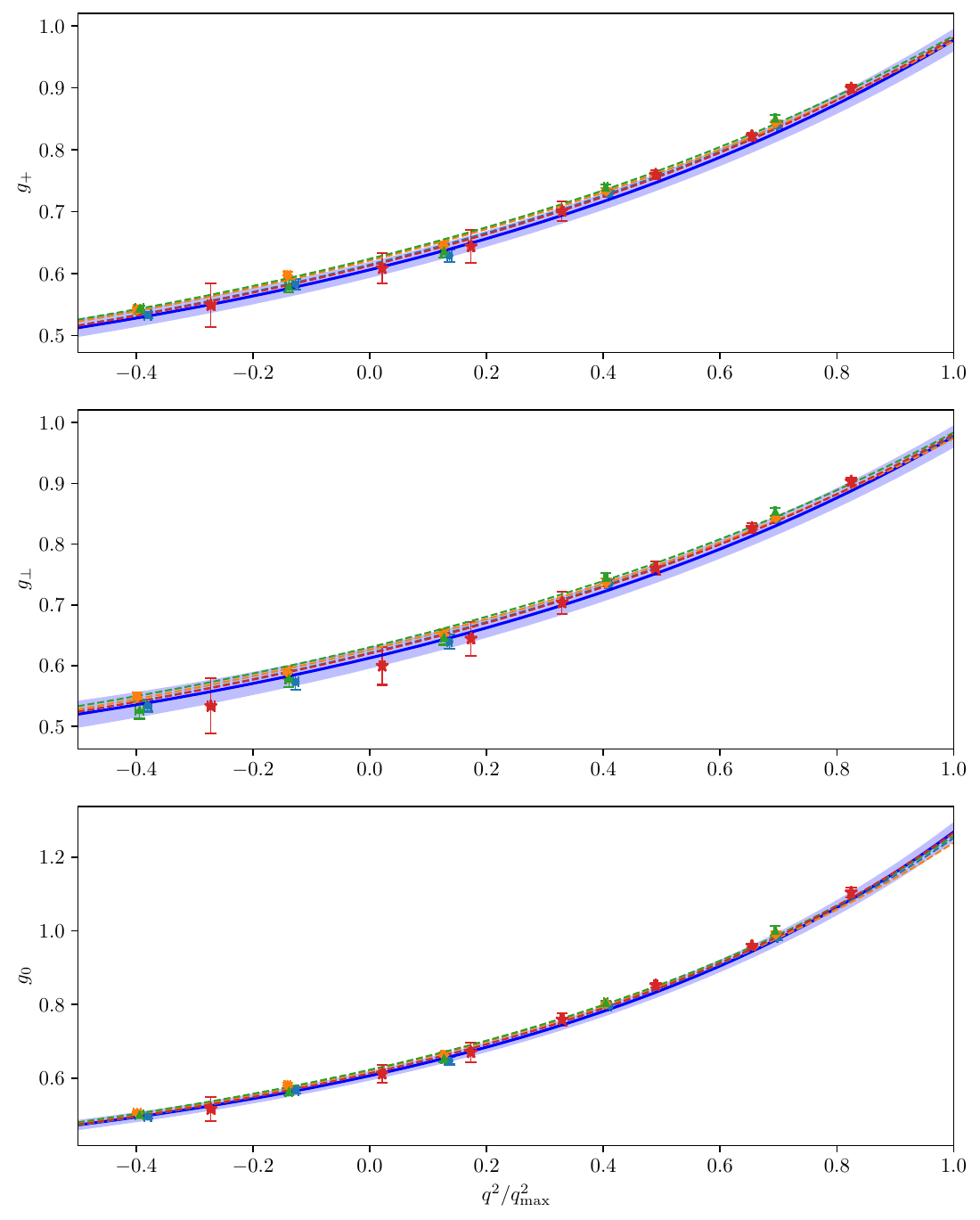}
    \caption{Like Fig.~\ref{fig:vector_extrap}, but for the axial-vector form factors.}
    \label{fig:axial_vector_extrap}
\end{figure}

\begin{figure}
    \centering
    \includegraphics[width=\linewidth]{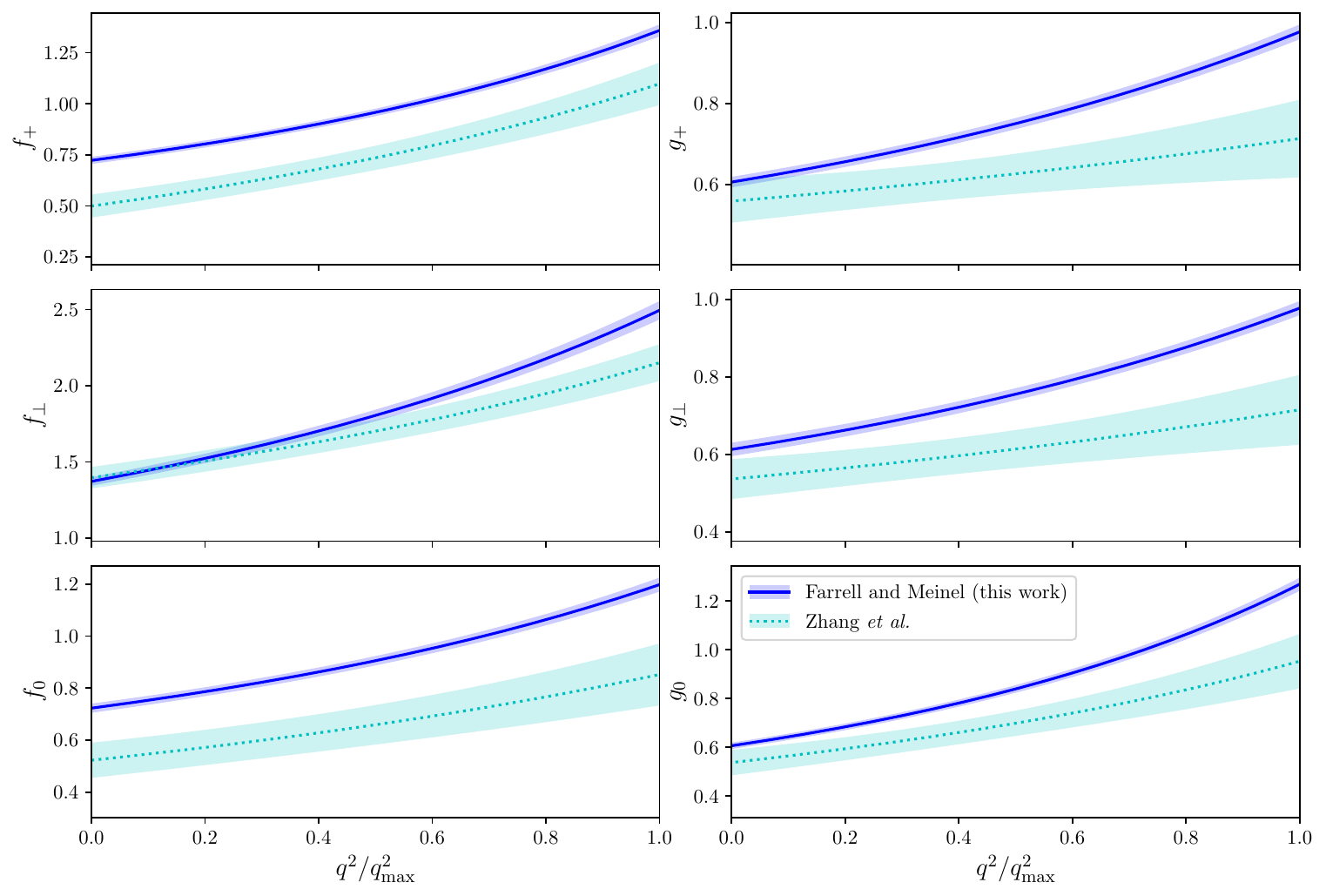}
    \caption{Comparison of the chiral/continuum-extrapolated form factors obtained in this work (solid lines) with the continuum-extrapolated form factors of Ref.~\cite{Zhang:2021oja} (dashed lines).}
    \label{fig:ZhangFFcompare}
\end{figure}

The central values and uncertainties of the parameters that are needed to evaluate the physical-limit form factors are listed in Table~\ref{tab:zfit_params}, and the correlation matrix is given in Table \ref{tab:corr} in the Appendix. Plots of the chiral-continuum extrapolations of the form factors are shown in Figs.~\ref{fig:vector_extrap} and \ref{fig:axial_vector_extrap} for the vector and axial-vector form factors, respectively. In addition, in Fig.~\ref{fig:ZhangFFcompare} we show a comparison of our results for the physical-limit form factors to those from Ref.~\cite{Zhang:2021oja}.

\FloatBarrier
\section{Decay-Rate Predictions}
\FloatBarrier

\begin{figure}
    \includegraphics[width=0.49\linewidth]{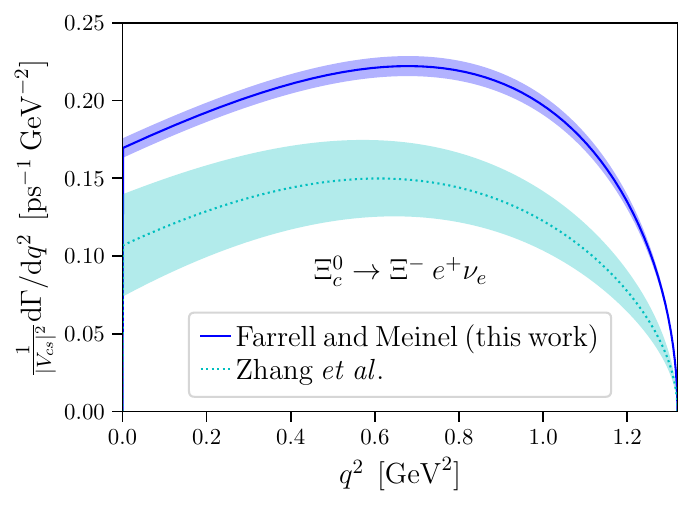}
    \hfill
    \includegraphics[width=0.49\linewidth]{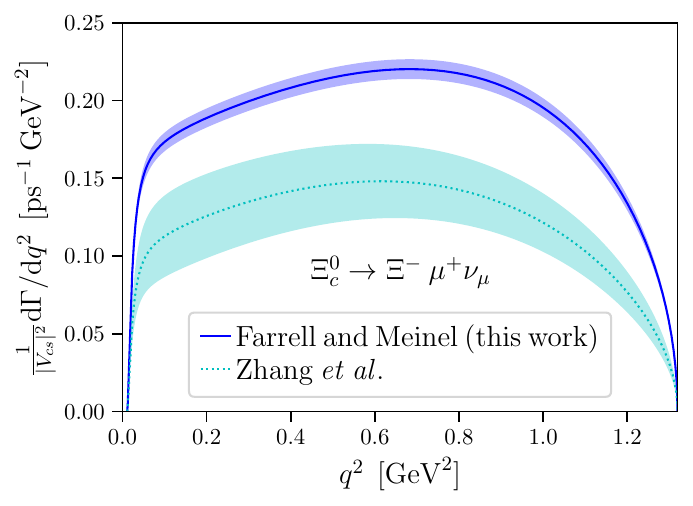}
    \caption{\label{fig:dGamma}Lattice predictions for the $\Xi_c^0 \to \Xi^- e^+ \nu_e$ and $\Xi_c^0 \to \Xi^- \mu^+ \nu_\mu$ differential decay rates in the Standard Model, without the $|V_{cs}|^2$ factor. The solid lines denote the predictions from this work, while the dashed lines are the prediction made using the form factors of Ref.~\cite{Zhang:2021oja}.}
\end{figure}

The Standard-Model expression for the $\Xi_c\to\Xi\ell\nu_\ell$ differential decay rate in terms of the helicity form factors is given, for example, in Eq.~(17) of Ref.~\cite{Meinel:2017ggx} (with the appropriate replacements of the CKM matrix element and baryon masses). When evaluating this expression, we take the hadron and lepton masses from experiment \cite{ParticleDataGroup:2024cfk}. Our predictions for the $\Xi_c^0 \to \Xi^- e^+ \nu_e$ and $\Xi_c^0 \to \Xi^- \mu^+ \nu_\mu$ differential decay rates divided by $|V_{cs}|^2$ are shown in Fig.~\ref{fig:dGamma}, in which we also included (our evaluations of) the predictions using the form factors of Ref.~\cite{Zhang:2021oja} for comparison. The corresponding figures for the $\Xi_c^+$ decays would differ only through small changes in the baryon masses and would look nearly identical. Our Standard-Model predictions for the integrated rates are
\begin{align}
\frac{1}{|V_{cs}|^2}\Gamma(\Xi_c^0 \to \Xi^- e^+ \nu_e) &= 0.2515(73)\text{ ps}^{-1},  \\
\frac{1}{|V_{cs}|^2}\Gamma(\Xi_c^0 \to \Xi^- \mu^+ \nu_\mu) &= 0.2437(71)\text{ ps}^{-1}, \\
\frac{1}{|V_{cs}|^2}\Gamma(\Xi_c^+ \to \Xi^0 e^+ \nu_e) &= 0.2549(74)\text{ ps}^{-1},  \\
\frac{1}{|V_{cs}|^2}\Gamma(\Xi_c^+ \to \Xi^0 \mu^+ \nu_\mu) &= 0.2471(72)\text{ ps}^{-1}.
\end{align}
The uncertainties given are the total uncertainties (including systematic uncertainties, as discussed in the previous section).

Taking the lifetime values $\tau_{\Xi^0_c}=150.4(2.8) \text{ fs}$, $\tau_{\Xi_c^+}=454(5)\text{ fs}$ from Ref.~\cite{ParticleDataGroup:2024cfk} and $|V_{cs}|=0.97345(20)$ from Ref.~\cite{UTfit:2022hsi}, we additionally obtain the following Standard-Model predictions for the branching fractions:
\begin{align}
\mathcal{B}(\Xi_c^0 \to \Xi^- e^+ \nu_e) &= 3.58(12)\:\%, \label{eq:BXic0eLattice} \\
\mathcal{B}(\Xi_c^0 \to \Xi^- \mu^+ \nu_\mu) &= 3.47(12)\:\%, \\
\mathcal{B}(\Xi_c^+ \to \Xi^0 e^+ \nu_e) &= 10.94(34)\:\%,\\
\mathcal{B}(\Xi_c^+ \to \Xi^0 \mu^+ \nu_\mu) &= 10.61(33)\:\%.
\end{align}

\FloatBarrier
\section{Conclusions}
\FloatBarrier

We have presented a new lattice calculation of the $\Xi_c \to \Xi$ vector and axial-vector form factors, leading to Standard-Model predictions of the $\Xi_c \to \Xi \ell^+\nu_\ell$ branching fractions with about 3\% total uncertainty. Our results for the form factors, and hence the branching fractions, are more precise and higher than those from the previous lattice calculation of Zhang {\it et al.} \cite{Zhang:2021oja}. While no chiral extrapolation was performed in Ref.~\cite{Zhang:2021oja}, we do not observe much pion-mass dependence in the form factors here, meaning that the lack of this extrapolation in Ref.~\cite{Zhang:2021oja} is unlikely to be the explanation for the discrepancy. Other possible origins include a statistical fluctuation, residual excited-state contamination, or discretization errors not removed by the continuum extrapolation (in particular, heavy-quark discretization errors \cite{El-Khadra:1996wdx}). Also note that the methods used for the renormalization of the $c\to s$ currents differ, and that the currents in Ref.~\cite{Zhang:2021oja} were not $O(a)$-improved.

Our Standard-Model result for $\mathcal{B}(\Xi_c^+ \to \Xi^0 e^+ \nu_e)$ is consistent with the experimental value of $(7\pm 4)\%$ reported by the PDG \cite{ParticleDataGroup:2024cfk} based on the measurement by CLEO \cite{CLEO:1994aud}. On the other hand, for the semileptonic decays of the neutral $\Xi_c^0$, our calculation predicts branching fractions in the Standard Model that are much higher than the more recent experimental results [compare Eq.~(\ref{eq:BXic0eLattice}) with Eqs.~(\ref{eq:BXic0eBelle}-\ref{eq:BXic0ePDG})], but reasonably close to the expectation from approximate $SU(3)$ flavor symmetry based on the experimental results for $\Lambda_c$ decays [see Eqs.~(\ref{eq:BXic0SU3LcL}) and (\ref{eq:BXic0SU3Lcn})]. Our Standard-Model prediction for $\mathcal{B}(\Xi_c^0 \to \Xi^- e^+ \nu_e)$ is higher than the 2024 PDG value  \cite{ParticleDataGroup:2024cfk} by a factor of approximately 3.4, and the significance of this discrepancy is $10.8\sigma$.

Recall from Sec.~\ref{sec:intro} that the experimental values for $\mathcal{B}(\Xi_c^0 \to \Xi^- e^+ \nu_e)$ are obtained through multiplying the measured ratio $\mathcal{B}(\Xi_c^0 \rightarrow \Xi^- e^+ \nu_{e})/\mathcal{B}(\Xi_c^0 \rightarrow \Xi^- \pi^+)$ by a separate measurement of $\mathcal{B}(\Xi_c^0 \rightarrow \Xi^- \pi^+)$. Recent global analyses of multiple charm-baryon nonleptonic decays using $SU(3)$ flavor symmetry indicate that the experimentally measured value of $\mathcal{B}(\Xi_c^0 \rightarrow \Xi^- \pi^+)$ may be underestimated \cite{Geng:2023pkr,Xing:2024nvg}. Therefore, new measurements of this normalization mode could help resolve the current puzzle.

\FloatBarrier
\section*{Acknowledgments}
\FloatBarrier

We thank Qi-An Zhang for providing bootstrap samples for the form-factor parameters from Ref.~\cite{Zhang:2021oja}, which we used to make the comparison plots. We thank Forrest Guyton for providing the $D^*$ masses that enter in our chiral-continuum extrapolation. We are grateful to the RBC and UKQCD Collaborations for making their gauge-field ensembles available. We acknowledge financial support by the U.S. Department of Energy, Office of Science, Office of High Energy Physics under Award Number DE-SC0009913.
This research used resources of the National Energy Research Scientific Computing Center (NERSC), a U.S.~Department of Energy Office of Science User Facility supported by Contract Number DE-AC02-05CH1123. This research also used resources at Purdue University RCAC and at the University of Texas TACC through the Extreme Science and Engineering Discovery Environment (XSEDE) \cite{XSEDE} and the Advanced Cyberinfrastructure Coordination Ecosystem: Services \& Support (ACCESS) program \cite{10.1145/3569951.3597559}, which are supported by U.S. National Science Foundation grants ACI-154856, 2138259, 2138286, 2138307, 2137603, and 2138296.  We acknowledge the use of Chroma \cite{Edwards:2004sx,Chroma}, QLUA \cite{QLUA}, MDWF \cite{MDWF}, and related USQCD software \cite{USQCD}.

\FloatBarrier
\section{Appendix}

\begin{figure}
    \centering
    \includegraphics[width=\linewidth]{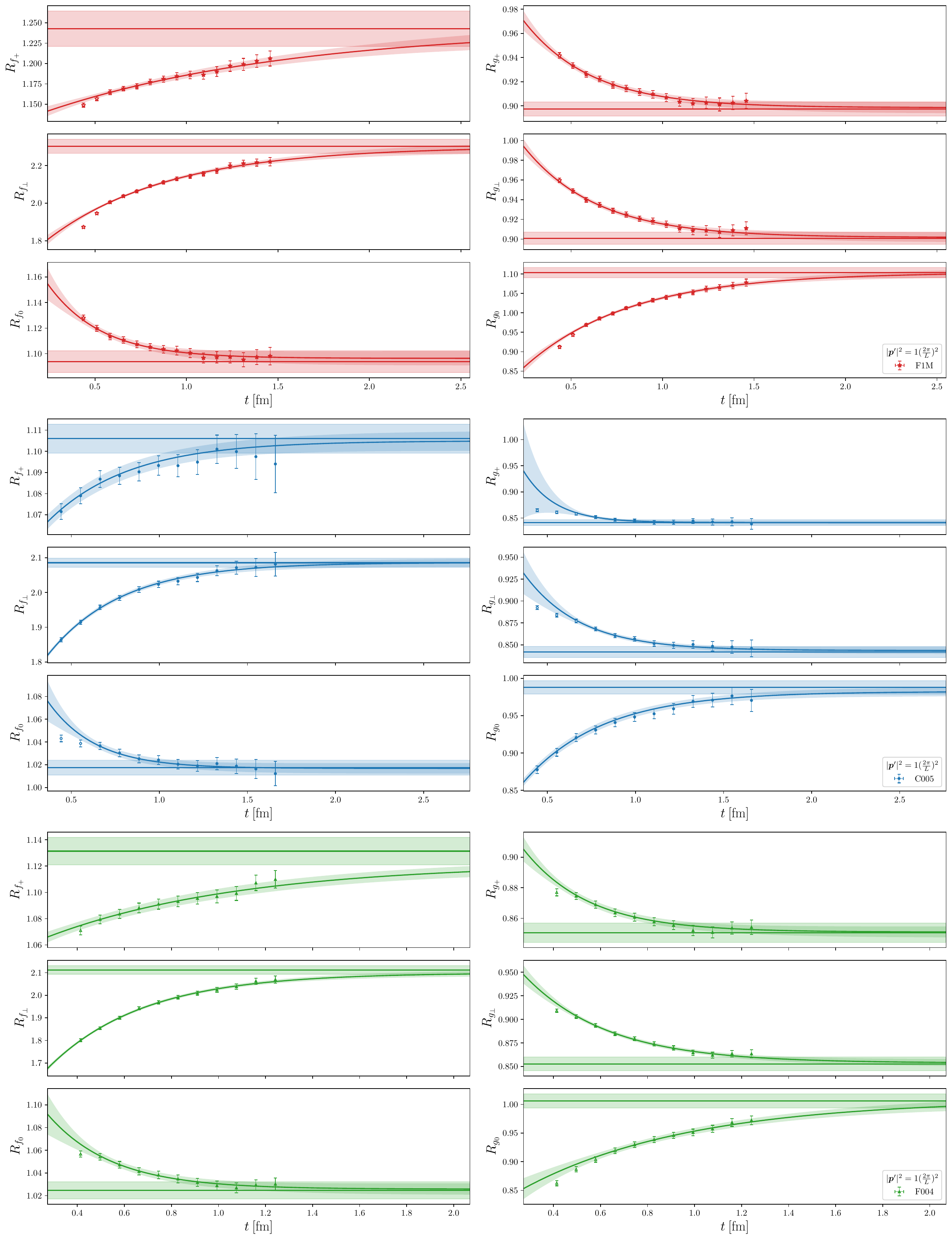}

    \caption{Like Fig.~\ref{fig:c01_ratios}, but for the other ensembles. As described in the main text, the ensembles $\rm{C005,C01,}$ and $\rm{F004}$ are fit simultaneously, while $\rm{F1M}$ is fit separately. }
    \label{fig:other_ensemble_ratios}
\end{figure}

This appendix contains sample plots of the ratio fits for the other ensembles (Fig.~\ref{fig:other_ensemble_ratios}) and the correlation matrix of the $z$-expansion parameters (Table \ref{tab:corr}).

\begin{turnpage}

\begin{table}
\hspace{-3em}
\begin{adjustbox}{width=1.3\textwidth}
\begin{tabular}{lccccccccccccccccccccc}
\hline\hline
 & $a_0^{f_{+}}$ & $a_1^{f_{+}}$ & $a_2^{f_{+}}$ & $a_3^{f_{+}}$ & $a_0^{f_{\perp}}$ & $a_1^{f_{\perp}}$ & $a_2^{f_{\perp}}$ & $a_3^{f_{\perp}}$ & $a_0^{f_{0}}$ & $a_1^{f_{0}}$ & $a_3^{f_{0}}$ & $a_0^{g_{+},g_{\perp}}$ & $a_1^{g_{+}}$ & $a_2^{g_{+}}$ & $a_3^{g_{+}}$ & $a_1^{g_{\perp}}$ & $a_2^{g_{\perp}}$ & $a_3^{g_{\perp}}$ & $a_0^{g_{0}}$ & $a_1^{g_{0}}$ & $a_3^{g_{0}}$ \\
\hline
$a_0^{f_{+}}$ & $ 1 $ & $-0.6314$ & $ \wm 0.3079$ & $-0.0825$ & $ \wm 0.5338$ & $-0.2758$ & $ \wm 0.0544$ & $-0.0077$ & $ \wm 0.7310$ & $-0.3175$ & $-0.0074$ & $ \wm 0.1727$ & $-0.0975$ & $ \wm 0.0217$ & $ \wm 0.0019$ & $-0.1181$ & $ \wm 0.0217$ & $-0.0059$ & $ \wm 0.2113$ & $-0.1565$ & $ \wm 0.0055$ \\
$a_1^{f_{+}}$ & $-0.6314$ & $ 1 $ & $-0.7098$ & $ \wm 0.3318$ & $-0.3223$ & $ \wm 0.3420$ & $-0.1009$ & $-0.0067$ & $-0.3117$ & $ \wm 0.5454$ & $ \wm 0.0260$ & $-0.0974$ & $ \wm 0.1644$ & $-0.0490$ & $ \wm 0.0089$ & $ \wm 0.1784$ & $-0.0468$ & $-0.0057$ & $-0.1571$ & $ \wm 0.2141$ & $-0.0023$ \\
$a_2^{f_{+}}$ & $ \wm 0.3079$ & $-0.7098$ & $ 1 $ & $-0.7487$ & $ \wm 0.0842$ & $-0.1813$ & $ \wm 0.1948$ & $ \wm 0.0193$ & $ \wm 0.1145$ & $-0.2660$ & $-0.0441$ & $ \wm 0.0331$ & $-0.0522$ & $ \wm 0.0686$ & $-0.0243$ & $-0.0812$ & $ \wm 0.1006$ & $ \wm 0.0093$ & $ \wm 0.0398$ & $-0.0614$ & $ \wm 0.0006$ \\
$a_3^{f_{+}}$ & $-0.0825$ & $ \wm 0.3318$ & $-0.7487$ & $ 1 $ & $-0.0111$ & $ \wm 0.0137$ & $-0.0075$ & $ \wm 0.0066$ & $-0.0030$ & $-0.0193$ & $ \wm 0.0628$ & $-0.0006$ & $-0.0048$ & $ \wm 0.0116$ & $-0.0059$ & $-0.0140$ & $ \wm 0.0209$ & $ \wm 0.0009$ & $-0.0034$ & $ \wm 0.0065$ & $ \wm 0.0096$ \\
$a_0^{f_{\perp}}$ & $ \wm 0.5338$ & $-0.3223$ & $ \wm 0.0842$ & $-0.0111$ & $ 1 $ & $-0.6629$ & $ \wm 0.2201$ & $-0.0338$ & $ \wm 0.4443$ & $-0.2165$ & $-0.0005$ & $ \wm 0.1274$ & $-0.1063$ & $-0.0027$ & $ \wm 0.0069$ & $-0.0946$ & $-0.0086$ & $ \wm 0.0029$ & $ \wm 0.1877$ & $-0.1881$ & $-0.0066$ \\
$a_1^{f_{\perp}}$ & $-0.2758$ & $ \wm 0.3420$ & $-0.1813$ & $ \wm 0.0137$ & $-0.6629$ & $ 1 $ & $-0.6965$ & $ \wm 0.2547$ & $-0.1929$ & $ \wm 0.2386$ & $-0.0037$ & $-0.1017$ & $ \wm 0.1396$ & $-0.0332$ & $ \wm 0.0079$ & $ \wm 0.1205$ & $-0.0318$ & $-0.0116$ & $-0.1514$ & $ \wm 0.1999$ & $ \wm 0.0125$ \\
$a_2^{f_{\perp}}$ & $ \wm 0.0544$ & $-0.1009$ & $ \wm 0.1948$ & $-0.0075$ & $ \wm 0.2201$ & $-0.6965$ & $ 1 $ & $-0.6589$ & $ \wm 0.0354$ & $-0.0700$ & $ \wm 0.0110$ & $ \wm 0.0322$ & $-0.0483$ & $ \wm 0.0767$ & $-0.0238$ & $-0.0580$ & $ \wm 0.1012$ & $ \wm 0.0102$ & $ \wm 0.0166$ & $-0.0364$ & $-0.0056$ \\
$a_3^{f_{\perp}}$ & $-0.0077$ & $-0.0067$ & $ \wm 0.0193$ & $ \wm 0.0066$ & $-0.0338$ & $ \wm 0.2547$ & $-0.6589$ & $ 1 $ & $-0.0105$ & $-0.0026$ & $ \wm 0.0045$ & $-0.0024$ & $ \wm 0.0083$ & $-0.0035$ & $-0.0057$ & $ \wm 0.0052$ & $ \wm 0.0016$ & $ \wm 0.0028$ & $-0.0001$ & $-0.0011$ & $-0.0044$ \\
$a_0^{f_{0}}$ & $ \wm 0.7310$ & $-0.3117$ & $ \wm 0.1145$ & $-0.0030$ & $ \wm 0.4443$ & $-0.1929$ & $ \wm 0.0354$ & $-0.0105$ & $ 1 $ & $-0.5787$ & $-0.0518$ & $ \wm 0.2072$ & $-0.1241$ & $ \wm 0.0326$ & $-0.0043$ & $-0.1089$ & $ \wm 0.0397$ & $-0.0065$ & $ \wm 0.1854$ & $-0.1147$ & $ \wm 0.0046$ \\
$a_1^{f_{0}}$ & $-0.3175$ & $ \wm 0.5454$ & $-0.2660$ & $-0.0193$ & $-0.2165$ & $ \wm 0.2386$ & $-0.0700$ & $-0.0026$ & $-0.5787$ & $ 1 $ & $ \wm 0.3544$ & $-0.1196$ & $ \wm 0.2106$ & $-0.0873$ & $ \wm 0.0198$ & $ \wm 0.1936$ & $-0.0742$ & $-0.0036$ & $-0.1195$ & $ \wm 0.1694$ & $-0.0096$ \\
$a_3^{f_{0}}$ & $-0.0074$ & $ \wm 0.0260$ & $-0.0441$ & $ \wm 0.0628$ & $-0.0005$ & $-0.0037$ & $ \wm 0.0110$ & $ \wm 0.0045$ & $-0.0518$ & $ \wm 0.3544$ & $ 1 $ & $-0.0018$ & $ \wm 0.0097$ & $-0.0082$ & $ \wm 0.0045$ & $ \wm 0.0164$ & $-0.0124$ & $ \wm 0.0030$ & $-0.0030$ & $ \wm 0.0066$ & $-0.0041$ \\
$a_0^{g_{+},g_{\perp}}$ & $ \wm 0.1727$ & $-0.0974$ & $ \wm 0.0331$ & $-0.0006$ & $ \wm 0.1274$ & $-0.1017$ & $ \wm 0.0322$ & $-0.0024$ & $ \wm 0.2072$ & $-0.1196$ & $-0.0018$ & $ 1 $ & $-0.4707$ & $ \wm 0.1499$ & $-0.0253$ & $-0.3231$ & $ \wm 0.0860$ & $-0.0212$ & $ \wm 0.7276$ & $-0.2438$ & $-0.0096$ \\
$a_1^{g_{+}}$ & $-0.0975$ & $ \wm 0.1644$ & $-0.0522$ & $-0.0048$ & $-0.1063$ & $ \wm 0.1396$ & $-0.0483$ & $ \wm 0.0083$ & $-0.1241$ & $ \wm 0.2106$ & $ \wm 0.0097$ & $-0.4707$ & $ 1 $ & $-0.6695$ & $ \wm 0.3733$ & $ \wm 0.5662$ & $-0.1735$ & $ \wm 0.0221$ & $-0.2180$ & $ \wm 0.3680$ & $-0.0126$ \\
$a_2^{g_{+}}$ & $ \wm 0.0217$ & $-0.0490$ & $ \wm 0.0686$ & $ \wm 0.0116$ & $-0.0027$ & $-0.0332$ & $ \wm 0.0767$ & $-0.0035$ & $ \wm 0.0326$ & $-0.0873$ & $-0.0082$ & $ \wm 0.1499$ & $-0.6695$ & $ 1 $ & $-0.8110$ & $-0.2985$ & $ \wm 0.2712$ & $-0.0268$ & $ \wm 0.0345$ & $-0.1075$ & $-0.0029$ \\
$a_3^{g_{+}}$ & $ \wm 0.0019$ & $ \wm 0.0089$ & $-0.0243$ & $-0.0059$ & $ \wm 0.0069$ & $ \wm 0.0079$ & $-0.0238$ & $-0.0057$ & $-0.0043$ & $ \wm 0.0198$ & $ \wm 0.0045$ & $-0.0253$ & $ \wm 0.3733$ & $-0.8110$ & $ 1 $ & $ \wm 0.0666$ & $-0.0955$ & $ \wm 0.0313$ & $ \wm 0.0271$ & $-0.0394$ & $ \wm 0.0611$ \\
$a_1^{g_{\perp}}$ & $-0.1181$ & $ \wm 0.1784$ & $-0.0812$ & $-0.0140$ & $-0.0946$ & $ \wm 0.1205$ & $-0.0580$ & $ \wm 0.0052$ & $-0.1089$ & $ \wm 0.1936$ & $ \wm 0.0164$ & $-0.3231$ & $ \wm 0.5662$ & $-0.2985$ & $ \wm 0.0666$ & $ 1 $ & $-0.6062$ & $ \wm 0.3079$ & $-0.2387$ & $ \wm 0.3044$ & $-0.0594$ \\
$a_2^{g_{\perp}}$ & $ \wm 0.0217$ & $-0.0468$ & $ \wm 0.1006$ & $ \wm 0.0209$ & $-0.0086$ & $-0.0318$ & $ \wm 0.1012$ & $ \wm 0.0016$ & $ \wm 0.0397$ & $-0.0742$ & $-0.0124$ & $ \wm 0.0860$ & $-0.1735$ & $ \wm 0.2712$ & $-0.0955$ & $-0.6062$ & $ 1 $ & $-0.6833$ & $ \wm 0.0115$ & $-0.0332$ & $ \wm 0.0612$ \\
$a_3^{g_{\perp}}$ & $-0.0059$ & $-0.0057$ & $ \wm 0.0093$ & $ \wm 0.0009$ & $ \wm 0.0029$ & $-0.0116$ & $ \wm 0.0102$ & $ \wm 0.0028$ & $-0.0065$ & $-0.0036$ & $ \wm 0.0030$ & $-0.0212$ & $ \wm 0.0221$ & $-0.0268$ & $ \wm 0.0313$ & $ \wm 0.3079$ & $-0.6833$ & $ 1 $ & $-0.0028$ & $-0.0185$ & $ \wm 0.0107$ \\
$a_0^{g_{0}}$ & $ \wm 0.2113$ & $-0.1571$ & $ \wm 0.0398$ & $-0.0034$ & $ \wm 0.1877$ & $-0.1514$ & $ \wm 0.0166$ & $-0.0001$ & $ \wm 0.1854$ & $-0.1195$ & $-0.0030$ & $ \wm 0.7276$ & $-0.2180$ & $ \wm 0.0345$ & $ \wm 0.0271$ & $-0.2387$ & $ \wm 0.0115$ & $-0.0028$ & $ 1 $ & $-0.6863$ & $-0.0719$ \\
$a_1^{g_{0}}$ & $-0.1565$ & $ \wm 0.2141$ & $-0.0614$ & $ \wm 0.0065$ & $-0.1881$ & $ \wm 0.1999$ & $-0.0364$ & $-0.0011$ & $-0.1147$ & $ \wm 0.1694$ & $ \wm 0.0066$ & $-0.2438$ & $ \wm 0.3680$ & $-0.1075$ & $-0.0394$ & $ \wm 0.3044$ & $-0.0332$ & $-0.0185$ & $-0.6863$ & $ 1 $ & $ \wm 0.3350$ \\
$a_3^{g_{0}}$ & $ \wm 0.0055$ & $-0.0023$ & $ \wm 0.0006$ & $ \wm 0.0096$ & $-0.0066$ & $ \wm 0.0125$ & $-0.0056$ & $-0.0044$ & $ \wm 0.0046$ & $-0.0096$ & $-0.0041$ & $-0.0096$ & $-0.0126$ & $-0.0029$ & $ \wm 0.0611$ & $-0.0594$ & $ \wm 0.0612$ & $ \wm 0.0107$ & $-0.0719$ & $ \wm 0.3350$ & $ 1 $ \\
\hline\hline
\end{tabular}
\end{adjustbox}
\hspace{-3em}
\caption{\label{tab:corr}Correlation matrix of the form factor parameters in Table~\ref{tab:zfit_params}. The covariance matrix is also provided as a machine-readable file in the supplemental material \cite{Supplemental}.}
\end{table}

\end{turnpage}

\FloatBarrier

\providecommand{\href}[2]{#2}\begingroup\raggedright\endgroup

\end{document}